\documentclass[english,aps,tightenlines,showpacs, showkeys, notitlepage]{revtex4-2}
\usepackage[LGR,T1]{fontenc}
\usepackage[latin9]{inputenc}
\usepackage{color}
\usepackage{babel}
\usepackage{amstext}
\usepackage{amssymb}
\usepackage{graphicx}
\usepackage{setspace}
\usepackage{esint}
\usepackage[unicode=true,pdfusetitle,
 bookmarks=true,bookmarksnumbered=false,bookmarksopen=false,
 breaklinks=false,pdfborder={0 0 0},pdfborderstyle={},backref=false,colorlinks=true]
 {hyperref}

\makeatletter

\DeclareRobustCommand{\greektext}{%
  \fontencoding{LGR}\selectfont\def\encodingdefault{LGR}}
\DeclareRobustCommand{\textgreek}[1]{\leavevmode{\greektext #1}}
\ProvideTextCommand{\~}{LGR}[1]{\char126#1}

\makeatother

\begin{document}
\title{Anisotropic cosmology in $q$-deformed entropic gravity}
\author{Salih Kibaro\u{g}lu$^{1}$}
\email{salihkibaroglu@maltepe.edu.tr}

\author{Mustafa Senay$^{2}$}
\email{msenay@bartin.edu.tr}

\date{\today}
\begin{abstract}
In this work, we examine the implications of $q$-deformed theory
on anisotropic Bianchi type-I cosmological model within the framework
of Verlinde's entropic gravity. The $q$-deformed theory, rooted in
quantum group structures, provides a natural generalization of classical
particle dynamics and offers a useful framework for exploring deviations
in standard cosmological models. By incorporating the principles of
entropic gravity, which conceptualize gravity as an emergent entropic
force, we derive the dynamical equations governing the Bianchi type-I
anisotropic model influenced by $q$-deformation. Finally, we analyze
the axisymmetric solution of Bianchi type-I model and these findings
highlight the potential impact of $q$-deformation on the dynamics
of the early universe and its subsequent evolution.
\end{abstract}
\affiliation{$^{1)}$Maltepe University, Faculty of Engineering and Natural Sciences,
34857, Istanbul, Türkiye}
\affiliation{$^{2)}$Department of Medical Services and Techniques, Vocational
School of Health Services, Bart\i n University, 74100, Bart\i n, Türkiye}
\maketitle

\section{Introduction}

General Relativity (GR) stands as one of the most successful theories
in modern physics. It provides a comprehensive framework for understanding
gravitational interactions, accurately describing the behavior of
massive objects and the large-scale structure of the universe. GR
has been extensively verified by experimental and observational data,
from the perihelion precession of Mercury to the deflection of light
by massive bodies and the recent detection of gravitational waves.
However, despite its successes, GR faces significant challenges when
applied to cosmological scales, particularly in explaining the accelerated
expansion of the universe and the nature of dark energy and dark matter.
These unresolved issues have motivated the development of modified
theories of gravity, which aim to extend or alter the framework of
GR to address these open questions. Modifications to GR often involve
introducing new degrees of freedom, additional fields, or altering
the geometric structure of spacetime itself. For instance, higher-dimensional
theories, scalar-tensor models, and deformations of spacetime symmetries
have been proposed as potential avenues to overcome the limitations
of GR (for more details, see reports \citep{Capozziello:2011Extended,Clifton:2011Modified,Nojiri:2017Modified}). 

In this context, $q$-deformed theories offer a promising alternative
by introducing deformation parameters into the algebraic structure
of physical systems \citep{Arik:1976Hilbert,Tsallis:1988Possible,Biedenharn:1989TheQuantum,Macfarlane:1989OnQAnalogs,Sun:1989TheQDeformed,Lavagno:2002Generalized,Jalalzadeh:2023Modified,Nojiri:2023Microscopic}.
While many modified gravity theories seek to address cosmological
phenomena by adjusting the geometric or scalar structure of spacetime,
the $q$-deformed theory introduces a different kind of modification:
it alters the algebraic structure of physical systems by introducing
a deformation parameter $q$, which leads to the formulation of $q$-deformed
algebras. By incorporating this deformation, the standard commutation
relations of quantum mechanics are generalized, enabling the description
of more complex physical behaviors that extend beyond conventional
frameworks.

These deformed algebraic structures provide a powerful framework for
the study of non-standard physical behavior and have a wide range
of applications across different areas of physics \citep{Wilczek:1990Fractional,Brito:2011q-Deformed,Bonatsos:1999Quantum,Senay:2018xaj,Kibaroglu:2018mnx,Kibaroglu:2019odt,Senay:2024Jeans}.
In condensed matter physics, $q$-deformed algebras are used to model
phenomena such as the fractional quantum Hall effect and the properties
of low-dimensional spin systems \citep{Wilczek:1990Fractional}. In
solid-state physics, they describe the thermal and electrical behavior
of materials with topological defects or exotic quasi-particle excitations
\citep{Brito:2011q-Deformed}. In nuclear physics, $q$-deformed statistics
provides better models for the thermodynamics of the quark-gluon plasma
and the spectral properties of hadrons \citep{Bonatsos:1999Quantum}.
More relevant to our discussion, in gravitational theory, $q$-deformed
algebras have been applied to explore the thermodynamics of black
holes and the entropy of the early universe. These applications offer
new perspectives on the fundamental nature of spacetime and potential
solutions to long-standing paradoxes in gravitational physics \citep{Senay:2018xaj,Kibaroglu:2018mnx,Kibaroglu:2019odt,Senay:2021Entropic,Senay:2021Heat,Senay:2024TheRoleMond}. 

The standard cosmological model, primarily based on the \textgreek{L}CDM
framework, has made significant achievements in explaining the large-scale
structure and evolution of the universe under the assumption of homogeneity
and isotropy. The Friedmann--Lemaître--Robertson--Walker (FLRW)
metric, which serves as the geometric foundation of this model, assumes
that the universe appears uniform when viewed on large scales, leading
to the well-supported predictions of cosmic expansion. This model
accounts for key phenomena such as the cosmic microwave background
(CMB), the abundance of light elements, and the large-scale distribution
of galaxies, all of which suggest that the universe has evolved in
a smooth, isotropic manner from an early hot and dense state \citep{Durrer:2015TheCosmic,Vakili:2012Classical}.
However, the assumption of perfect isotropy may be oversimplified,
particularly in the early universe. Motivated by potential deviations
from homogeneity at the very beginning of cosmic evolution, anisotropic
cosmological models have been proposed to capture the more complex
dynamics that could have existed before inflation. Recent observational
data, including findings from the Wilkinson Microwave Anisotropy Probe
(WMAP) \citep{WMAP:2003,WMAP:2006,WMAP:2008}, reveals the necessity
of extending beyond the standard isotropic and homogeneous cosmological
paradigm. The inflationary framework, strongly supported by observational
evidence, effectively demonstrates how an initially anisotropic and
inhomogeneous universe can evolve toward the isotropic and homogeneous
FLRW geometry observed today. Within this context, Bianchi-type cosmological
models serve as a comprehensive extension of standard models, preserving
spatial homogeneity while introducing anisotropic features. In recent
decades, these models have garnered significant interest in observational
cosmology, particularly in light of WMAP data, which suggest that
the standard cosmological model, incorporating a positive cosmological
constant, closely resembles Bianchi-type morphology \citep{Jaffe:2005Evidence,Jaffe:2006Bianchi,Jaffe:2006Fast,Campanelli:2006Ellipsoidal,Campanelli:2007Cosmic}.
Additionally, extensive studies of Bianchi cosmological models have
been conducted from diverse perspectives, offering critical insights
into the nature of primordial fluctuations and the large-scale structure
of the universe. These investigations present promising pathways for
addressing unresolved questions in cosmology \citep{Vollick:2003AnisotropicBI,Ellis:2006TheBianchiModels,Saha:2006Anisotropic,Gumrukcuoglu:2007Inflationary,Rodrigues:2008Evolution,Akarsu:2010Bianchi,Sharif:2011Anisotropic,Harko:2014Bianchi,Muller:2018Anisotropic,Kumar:2021Anisotropic,Costantini:2022AReconstruction,Nojiri:2022Formalizing,Parnovsky:2023TheBigBang,Kibaroglu:2025Anisotropic}.

In this study, we examine $q$-deformed entropic gravity model in
the context of the anisotropic evolution of the universe. Incorporating
anisotropic cosmologies within the framework of $q$-deformed entropic
gravity models presents a promising avenue for addressing the complexities
inherent in non-isotropic and non-homogeneous cosmological scenarios.
This integration introduces a deformation parameter, which allows
for more flexible modeling of gravitational and thermodynamic processes.
It potentially offers new insights into the behavior of cosmic anisotropies,
dark energy, and the evolution of the universe under conditions not
fully accounted for by classical theories. By doing so, it may open
the door to novel predictions about the early universe's behavior,
the role of anisotropies in large-scale structure formation, and the
influence of quantum statistical effects on the evolution of cosmological
models. 

The paper is organized as follows: in Section \ref{sec:Deformed-fermion-statistics},
we present the deformed fermion statistics, laying the groundwork
for the $q$-deformation formalism. Section \ref{sec:q-Deformed-Einstein-Equations}
introduces the $q$-deformed Einstein equations, exploring the impact
of deformation in the gravitational sector. In Section \ref{sec:Exploring-Anisotropic-Cosmology},
we investigate anisotropic cosmology, with a focus on axisymmetric
solutions, and analyze how $q$-deformation modifies the standard
cosmological models. Finally, the paper concludes with a summary of
the results and potential future directions in Section \ref{sec:Conclusion}.

\section{Deformed fermion statistics\label{sec:Deformed-fermion-statistics}}

In this section, we present the fundamental quantum algebraic and
thermodynamic properties of the $q$-deformed fermion gas model, which
provides an innovative approach to understanding quasiparticle systems
by extending beyond the Pauli exclusion principle. This algebraic
framework was introduced in Refs. \citep{Parthasarathy:1991A-Q-Analog,Viswanathan:1992Generalized,Chaichian:1993Statistics},
and its high- and low-temperature thermodynamic properties have been
examined in Refs. \citep{Algin:2012High,Algin:2016Fermionic,Algin:2016General}.
The algebra is governed by the following deformed anticommutation
relations:
\begin{equation}
ff^{*}+qf^{*}f=1,
\end{equation}
\begin{equation}
\left[\hat{N},f^{*}\right]=f^{*},\,\,\,\,\left[\hat{N},f\right]=-f,
\end{equation}
where $f$ and $f^{*}$ represent the deformed fermionic annihilation
and creation operators, respectively, while $\hat{N}$ denotes the
fermionic number operator. The deformation parameter $q$ is a positive
real number within the range $0<q<1$. The fermionic number operator
spectrum can be expressed as:

\begin{equation}
\left[n\right]=\frac{1-\left(-1\right)^{n}q^{n}}{1+q}.
\end{equation}
To investigate the thermodynamic properties of the deformed fermion
gas model, the fermionic Jackson derivative operator can be employed
as a substitute for the conventional derivative operator

\begin{equation}
D_{x}^{q}f(x)=\frac{1}{x}\left[\frac{f\left(x\right)-f\left(-qx\right)}{1+q}\right],
\end{equation}
applicable to any function $f(x)$. The average occupation number
for the deformed fermion gas model is given by:
\begin{equation}
n_{i}=\frac{1}{|\ln q|}\left|\ln\left(\frac{|1-ze^{\beta\epsilon_{i}}|}{1+qze^{\beta\epsilon_{i}}}\right)\right|,\label{eq: Tot_num_particles}
\end{equation}
where $\epsilon_{i}$ is the kinetic energy of a particle in the single-particle
energy state, $\beta=1/k_{B}T$, with $k_{B}$ being the Boltzmann
constant, and $T$ the temperature of the system. The fugacity $z=\exp(\mu/k_{B}T)$,
where $\mu$ denotes the chemical potential, retains its standard
form.

By utilizing Eq.(\ref{eq: Tot_num_particles}), the total number of
particles $N$ and the energy of the system $U$ can be correspondingly
expressed as:

\begin{equation}
N=\sum_{i}\frac{1}{|\ln q|}\left|\ln\left(\frac{|1-ze^{\beta\epsilon_{i}}|}{1+qze^{\beta\epsilon_{i}}}\right)\right|,
\end{equation}
\begin{equation}
U=\sum_{i}\frac{\epsilon_{i}}{|\ln q|}\left|\ln\left(\frac{|1-ze^{\beta\epsilon_{i}}|}{1+qze^{\beta\epsilon_{i}}}\right)\right|.
\end{equation}
In the regime of large volumes and particle numbers, the summations
over quantum states can be replaced by integrals. Using the density
of states $\left(V/2\pi^{2}\right)\left(2m/\hbar^{2}\right)^{3/2}\epsilon^{1/2}$,
the following relations can be derived:

\begin{equation}
\frac{N}{V}=\frac{1}{\lambda^{3}}f_{3/2}\left(z,q\right),
\end{equation}
\begin{equation}
\frac{U}{V}=\frac{3}{2}\frac{k_{B}T}{\lambda^{3}}f_{5/2}\left(z,q\right),
\end{equation}
where $V$ represents the volume of the system, $\lambda=h/\left(2\pi mk_{B}T\right)^{1/2}$
is the thermal wavelength, and the generalized Fermi-Dirac function
$f_{n}(z,q)$ is defined as

\begin{eqnarray}
f_{n}\left(z,q\right) & = & \frac{1}{\Gamma(n)}\int_{0}^{\infty}\frac{x^{n-1}}{|\ln q|}\left|\ln\left(\frac{|1-ze^{-x}|}{1+qze^{-x}}\right)\right|\text{d}x,
\end{eqnarray}
or equivalently expressed as a series:

\begin{eqnarray}
f_{n}\left(z,q\right) & = & \frac{1}{|\ln q|}\left[\sum_{l=1}^{\infty}\left(-1\right)^{l-1}\frac{\left(zq\right)^{l}}{l^{n+1}}-\sum_{l=1}^{\infty}\frac{z^{l}}{l^{n+1}}\right],
\end{eqnarray}
where $x=\beta\epsilon$. From the thermodynamic relation $F=\mu N-PV$
(here $P=2U/3V$ is the pressure of the system), the Helmholtz free
energy can be expressed as

\begin{equation}
F=\frac{k_{B}TV}{\lambda^{3}}\left[f_{3/2}\left(z,q\right)\ln z-f_{5/2}\left(z,q\right)\right].
\end{equation}
The entropy function of the model can be obtained using the relation
$S=\left(U-F\right)/T$, yielding: 
\begin{equation}
S=\frac{k_{B}V}{\lambda^{3}}\left[\frac{5}{2}f_{5/2}\left(z,q\right)-f_{3/2}\left(z,q\right)\ln z\right].\label{eq: Entropy_1}
\end{equation}
By assuming that the kinetic energy of a single particle is $E=k_{B}T$,
Eq.(\ref{eq: Entropy_1}) can be reformulated as 
\begin{equation}
S=\frac{\left(2\pi m\right)^{3/2}}{Th^{3}}E^{5/2}\tilde{F}\left(z,q\right),\label{eq: Entropy_deformed}
\end{equation}
where the following expression, which encodes the $q$-deformation
contribution, has been introduced for simplicity:

\begin{equation}
\tilde{F}\left(z,q\right)=\frac{5}{2}f_{5/2}\left(z,q\right)-f_{3/2}\left(z,q\right)\ln z.
\end{equation}
Thus, we demonstrated key thermodynamic properties, such as particle
number density, internal energy, Helmholtz free energy, and entropy,
illustrating how deformation influences the system. Building on this
analysis, the next section delves into the implications of this deformed
system within the gravitational sector, examining how $q$-deformation
influences gravitational dynamics and the interplay between quantum
statistical mechanics and gravitational fields. This transition allows
us to explore potential applications of the deformed fermion model
in cosmological and astrophysical contexts.

\section{$q$-Deformed Einstein Equations\label{sec:q-Deformed-Einstein-Equations}}

In this section, we investigate the influence of $q$-deformed fermions
on Einstein's field equations within the framework proposed by Verlinde
\citep{Verlinde:2010hp,Verlinde:2016toy}. Verlinde\textquoteright s
entropic gravity approach suggests that gravity is not a fundamental
force but an emergent phenomenon driven by changes in entropy. According
to Verlinde, when a test particle moves toward a holographic screen,
an entropic force arises, which can be described by the following
relation:
\begin{equation}
\mathcal{F}=T\frac{\Delta S}{\Delta x},
\end{equation}
where $\Delta S$ represents the change in entropy on the holographic
screen, and $\Delta x$ denotes the displacement of the particle from
the screen. The holographic screen is interpreted as the maximal information
storage surface, with the total number of bits $N$ proportional to
the area $A$ of the screen. The number of bits is given by:
\begin{equation}
N=\frac{A}{G\hbar},
\end{equation}
where the speed of light $c=1$, and $G$ and $\hbar$ are the gravitational
constant and reduced Planck constant, respectively.

When the entropic force counterbalances the force driving entropy
increase, the system reaches thermodynamic equilibrium, and the total
entropy remains constant. This equilibrium condition is expressed
as:

\begin{equation}
\frac{d}{dx^{a}}S\left(E,x^{a}\right)=0,
\end{equation}
Expanding this condition yields:

\begin{equation}
\frac{\partial S}{\partial E}\frac{\partial E}{\partial x^{a}}+\frac{\partial S}{\partial x^{a}}=0,\label{eq: Entropy_Energy_Relation}
\end{equation}
where $\frac{\partial E}{\partial x^{a}}=-F_{a}$ and $\frac{\partial S}{\partial x^{a}}=\nabla_{a}S$.
Substituting Eq.(\ref{eq: Entropy_deformed}) into Eq.(\ref{eq: Entropy_Energy_Relation})
leads to the expression for the deformed temperature, given by:
\begin{equation}
T=\frac{5V}{8\sqrt{\pi}}\frac{\left(2mE\right)^{3/2}}{h^{2}}\tilde{F}\left(z,q\right)e^{\phi}N^{a}\nabla_{a}\phi,
\end{equation}
where $F_{a}=me^{\phi}\nabla_{a}\phi$ and $\nabla_{a}S=-2\pi mN_{a}/\hbar$
have been used in the derivation. Here, $e^{\phi}$ represents the
redshift factor and $N^{a}$ is the unit outward normal vector. The
above expression for the temperature can be re-expressed as:

\begin{equation}
T=\alpha\left(z,q\right)T_{U},
\end{equation}
where $T_{U}=\frac{\hbar}{2\pi}e^{\phi}N^{a}\nabla_{a}\phi$ is the
Unruh temperature \citep{Unruh:1976Notes,Verlinde:2010hp} and $\alpha(z,q)$
encapsulates the effects of the $q$-deformed fermion system, defined
as:

\begin{equation}
\alpha\left(z,q\right)=\frac{5V\left(2\pi mE\right)^{3/2}}{2h^{3}}\tilde{F}\left(z,q\right).
\end{equation}
Using the mass-energy relation $M=N\frac{T}{2}$, the total mass can
be written as 
\begin{equation}
M=\frac{\alpha\left(z,q\right)}{4\pi G}\int_{\mathcal{S}}e^{\phi}\nabla\phi dA,\label{eq: Total_Mass}
\end{equation}
where $\mathcal{S}$ denotes the holographic screen. The integral
on the right-hand side relates to the modified Komar mass. Thus, Eq.(\ref{eq: Total_Mass})
can be viewed as a modified form of Gauss's law in general relativity.
Employing Stokes' theorem and following the approach outlined in Ref.\citep{Senay:2018xaj,Kibaroglu:2018mnx},
Eq.(\ref{eq: Total_Mass}) can be re-expressed in terms of the Killing
vector $\xi^{a}$ and the Ricci curvature tensor $R_{ab}$ as:
\begin{equation}
M=\frac{\alpha\left(z,q\right)}{4\pi G}\int_{\Sigma}R_{ab}n^{a}\xi^{b}d\Sigma.\label{eq: Total_Mass_2}
\end{equation}
Here, $\Sigma$ denotes the holographic screen. Furthermore, the Komar
mass can also be defined in terms of the stress-energy tensor $T_{ab}$
\citep{Wald:2010GeneralRelativity} as 
\begin{equation}
\mathcal{M}=2\int_{\Sigma}\left(T_{ab}-\frac{1}{2}g_{ab}T-\frac{\Lambda}{8\pi G}g_{ab}\right)n^{a}\xi^{b}d\Sigma,\label{eq: Komar_Mass}
\end{equation}
where $g_{ab}$ is metric tensor and $\Lambda$ is a constant that
can be associated with the cosmological constant. Assuming that the
stress-energy tensor $T_{\,\,b}^{a}=\text{diag}\left(\rho,-P,-P,-P\right)$
exhibits the characteristics of a perfect fluid together with the
energy density $\rho\left(t\right)$ and pressure $P\left(t\right)$.
By equating Eq.(\ref{eq: Total_Mass_2}) with Eq.(\ref{eq: Komar_Mass}),
we arrive at the following relation:

\begin{equation}
\alpha\left(z,q\right)R_{ab}=8\pi G\left(T_{ab}-\frac{1}{2}g_{ab}T-\frac{\Lambda}{8\pi G}g_{ab}\right).
\end{equation}
Taking the trace of the above equation, we obtain:

\begin{equation}
R_{ab}-\frac{1}{2}g_{ab}R-\tilde{\Lambda}g_{ab}=8\pi\tilde{G}T_{ab},\label{eq: EFE}
\end{equation}
where $\tilde{G}=G/\alpha\left(z,q\right)$ and $\tilde{\Lambda}=\Lambda/\alpha\left(z,q\right)$
can be interpreted as an effective gravitational constant and effective
cosmological constant, respectively. Thus, within the framework of
Verlinde's entropic gravity, we derive a modification of Einstein's
field equations as a result of the $q$-deformed fermion theory. This
demonstrates that the presence of $q$-deformation induces significant
modifications to the underlying gravitational dynamics. 

\section{Exploring Anisotropic Cosmology\label{sec:Exploring-Anisotropic-Cosmology}}

We now turn our attention to the investigation of a homogeneous and
anisotropic spacetime governed by the Bianchi type-I metric, which
constitutes the simplest extension of the spatially flat FLRW metric.
This class of metrics is distinguished by its spatial homogeneity
while permitting anisotropy in different spatial directions. Formally,
the Bianchi type-I metric is expressed as:
\begin{equation}
\text{d}s^{2}=\text{d}t^{2}-A\left(t\right)^{2}\text{d}x^{2}-B\left(t\right)^{2}\text{d}y^{2}-C\left(t\right)^{2}\text{d}z^{2},\label{eq: metric_anisotropic}
\end{equation}
where $A\left(t\right)$, $B\left(t\right)$ and $C\left(t\right)$
are the scale factors associated with the $x$, $y$ and $z$ axes,
respectively. These scale factors encapsulate the anisotropic expansion
or contraction of the universe along each spatial direction, distinguishing
the Bianchi type-I metric from the isotropic FLRW metric, which assumes
a single scale factor. Additionally, if we set $A\left(t\right)=B\left(t\right)=C\left(t\right)$,
without loss of generality the metric simplifies to its corresponding
isotropic FLRW form. The average expansion scale factor can be defined
as $a\left(t\right)=\left(ABC\right)^{1/3}$, and the corresponding
spatial volume as $V=ABC$ for this model. This model also assumes
spatial flatness, consistent with the widely accepted understanding
that the universe is spatially flat or nearly flat. 

The anisotropic behavior of the universe's expansion is effectively
described by introducing directional Hubble parameters corresponding
to the principal $x$, $y$, and $z$ axes;
\begin{equation}
H_{x}\left(t\right)=\frac{\dot{A}}{A},\,\,\,\,\,\,\,\,\,\,\,\,\,H_{y}\left(t\right)=\frac{\dot{B}}{B},\,\,\,\,\,\,\,\,\,\,\,\,\,H_{z}\left(t\right)=\frac{\dot{C}}{C},
\end{equation}
Additionally, the mean Hubble parameter is defined to represent the
overall expansion rate, providing a comprehensive characterization
of the anisotropy in the cosmological model,

\begin{equation}
H\left(t\right)=\left(1/3\right)\left(H_{x}+H_{y}+H_{z}\right).
\end{equation}
Moreover, the anisotropy parameter of the expansion is defined as
\begin{equation}
\Delta=\frac{1}{3}\sum_{i=1}^{3}\left(\frac{H_{i}-H}{H}\right)^{2},
\end{equation}
where $i=1,2,3$ and $H_{i}$ represent the directional Hubble parameters
along the $x$, $y$ and $z$ axes, respectively. The parameter $\Delta$
measures the degree of anisotropy in the expansion of the universe.
When $\Delta=0$, it indicates an isotropic expansion, meaning the
expansion rates along the $x$, $y$ and $z$-axes are identical,
thereby corresponding to the standard FLRW cosmology. Any non-zero
value of $\Delta$ reflects anisotropy, indicating differing expansion
rates along the principal axes. In this context, we also want to note
that the energy-momentum tensor can be expressed as

\begin{equation}
T_{\,\,b}^{a}=diag(\rho,-P_{x},-P_{y},-P_{z})
\end{equation}
where $\rho$$\left(t\right)$ denotes the energy density, and $P_{x}$$\left(t\right)$,
$P_{y}\left(t\right)$ and $P_{z}\left(t\right)$ represent the pressures
along the $x$, $y$ and $z$ directions, respectively.

Following the discussion of the anisotropic nature of the universe,
we now proceed to investigate the corresponding Friedmann equations
for this cosmological model. By inserting the Bianchi type-I metric
(\ref{eq: metric_anisotropic}) into the gravitational field equations
(\ref{eq: EFE}) and solving for the components of the Einstein tensor,
we can derive the corresponding modified Friedmann equations that
govern the dynamics of an anisotropic universe. In the following,
we will derive these equations step by step, beginning with the calculation
of the $\left\{ tt\right\} $-component of the field equations (\ref{eq: EFE}),
yields a modified first Friedmann equation:

\begin{equation}
\frac{1}{3}\left(H_{x}H_{y}+H_{x}H_{z}+H_{y}H_{z}\right)=\frac{8\pi\tilde{G}}{3}\rho+\frac{\tilde{\Lambda}}{3}.\label{eq: Friedmann_t}
\end{equation}
Next, we examine the $\left\{ xx\right\} $, $\left\{ yy\right\} $,
and $\left\{ zz\right\} $-components of the Einstein field equations,
which are given by:

\begin{equation}
H_{y}H_{z}+\frac{\ddot{B}}{B}+\frac{\ddot{C}}{C}=-8\pi\tilde{G}P_{x}+\tilde{\Lambda},\label{eq: Friedmann_x}
\end{equation}

\begin{equation}
H_{x}H_{z}+\frac{\ddot{A}}{A}+\frac{\ddot{C}}{C}=-8\pi\tilde{G}P_{y}+\tilde{\Lambda},\label{eq: Friedmann_y}
\end{equation}

\begin{equation}
H_{x}H_{y}+\frac{\ddot{A}}{A}+\frac{\ddot{B}}{B}=-8\pi\tilde{G}P_{z}+\tilde{\Lambda},\label{eq: Friedmann_z}
\end{equation}
By combining these equations, we obtain the following equation, which
describes the averaged acceleration of the universe:

\begin{equation}
\left(\frac{\ddot{A}}{A}+\frac{\ddot{B}}{B}+\frac{\ddot{C}}{C}\right)=-4\pi\tilde{G}\left(\rho+P_{x}+P_{y}+P_{z}\right),\label{eq: Friedmann_acc}
\end{equation}
This equation mirrors the structure of the standard acceleration equation
but with a generalized right-hand side that includes the contributions
from the anisotropic pressures along all three spatial axes. Additionally,
the continuity equation, which governs the conservation of energy,
generalizes to:

\begin{equation}
\dot{\rho}=-\left[H_{x}\left(\rho+P_{x}\right)+H_{y}\left(\rho+P_{y}\right)+H_{z}\left(\rho+P_{z}\right)\right],
\end{equation}
This extended continuity equation accounts for the anisotropic nature
of the energy and pressure distribution, generalizing the standard
form of the continuity equation $\dot{\rho}=-3H\left(\rho+P\right)$
by incorporating the directional Hubble parameters for each spatial
direction. 

\subsection{Axisymmetric solution}

The standard Bianchi type-I metric leads to highly complicated field
equations Eqs.(\ref{eq: Friedmann_t})-(\ref{eq: Friedmann_z}), posing
significant challenges to analyze this model. To streamline the mathematical
framework of the model under consideration, this study adopts the
axisymmetric Bianchi type-I metric. This choice mitigates the complexity
of the equations while preserving the key characteristics of anisotropic
cosmology, thereby facilitating a more manageable analysis and enabling
the derivation of viable cosmological solutions within the theoretical
framework. Specifically, we assume $A\left(t\right)=B\left(t\right)$
in the metric (\ref{eq: metric_anisotropic}), representing a subclass
of the Bianchi type-I model that exhibits axisymmetry about the $z$-axis.
This metric, widely acknowledged as the most general plane-symmetric
line element \citep{Taub:1951Empty}, is derived by specifying the
$xy$-plane as the plane of symmetry which is described by the metric
in Cartesian coordinates:

\begin{equation}
\text{d}s^{2}=-\text{d}t^{2}+A\left(t\right)^{2}\text{d}x^{2}+A\left(t\right)^{2}\text{d}y^{2}+C\left(t\right)^{2}\text{d}z^{2}.\label{eq: metric_axisymmetric}
\end{equation}
The adoption of the chosen Bianchi-I metric structure is justified
by its alignment with azimuthally symmetric physical configurations,
as extensively discussed in \citep{Barrow:2006Cosmologies,Berera:2004TheEccentric,Campanelli:2006Ellipsoidal}.
In theoretical and observational studies, this spacetime framework
is frequently identified as the \textquotedbl Ellipsoidal Universe\textquotedbl{}
or the \textquotedbl Eccentric Universe\textquotedbl{} \citep{Berera:2004TheEccentric,Campanelli:2009AModel,Deliduman:2024ellipsoidal}.
Moreover, by setting $A\left(t\right)=C\left(t\right)$ in the axisymmetric
metric (\ref{eq: metric_axisymmetric}), and applying a coordinate
rescaling to $x$, $y$ and $z$, the metric seamlessly transitions
into its isotropic FLRW counterpart, retaining its essential cosmological
features.

In this framework, we consider a vacuum solution, where the energy-momentum
tensor is assumed to be zero, i.e., $T_{\,\,b}^{a}=0$, corresponding
to a matterless scenario. This assumption implies that there is no
contribution from matter, radiation, or any other form of energy density.
Specifically, the modified Friedmann equations governing the evolution
of the anisotropic universe now take the following forms:

\begin{equation}
\frac{1}{3}\left(H_{xy}+2H_{xy}H_{z}\right)=\frac{\tilde{\Lambda}}{3},\label{eq: Friedmann_ax_t}
\end{equation}

\begin{equation}
H_{xy}H_{z}+\frac{\ddot{A}}{A}+\frac{\ddot{C}}{C}=\tilde{\Lambda},\label{eq: Friedmann_ax_xy}
\end{equation}

\begin{equation}
H_{xy}^{2}+2\frac{\ddot{A}}{A}=\tilde{\Lambda},\label{eq: Friedmann_ax_z}
\end{equation}
where $H_{xy}\left(t\right)=\frac{\dot{A}}{A}$ is the Hubble parameter
related with $x$ and $y$ directions. In this scenario, the universe
evolves purely under the influence of the effective cosmological constant,
with no additional matter or radiation contributing to the dynamics.
We now proceed to solve Eq.(\ref{eq: Friedmann_ax_z}), from which
the scale factor $A\left(t\right)$, the Hubble parameter $H_{xy}\left(t\right)$
and the effective equation of state parameter $w_{xy}\left(t\right)$
for $x$ and $y$ directions are obtained as follows:

\begin{equation}
A\left(t\right)=\frac{2^{-\frac{5}{3}}3^{\frac{1}{3}}}{e^{\sqrt{3\tilde{\Lambda}}t}}\left[\frac{1}{\sqrt{\tilde{\Lambda}}}\left(C_{1}e^{\sqrt{3\tilde{\Lambda}}t}+C_{2}\right)e^{\sqrt{3\tilde{\Lambda}}t}\right]^{\frac{2}{3}},\label{eq: A_ax}
\end{equation}

\begin{equation}
H_{xy}\left(t\right)=\sqrt{\frac{\tilde{\Lambda}}{3}}\left(\frac{C_{1}e^{\sqrt{3\tilde{\Lambda}}t}-C_{2}}{C_{1}e^{\sqrt{3\tilde{\Lambda}}t}+C_{2}}\right),\label{eq: H_xy_ax}
\end{equation}

\begin{equation}
w_{xy}\left(t\right)=-\left(\frac{C_{1}e^{\sqrt{3\tilde{\Lambda}}t}+C_{2}}{C_{1}e^{\sqrt{3\tilde{\Lambda}}t}-C_{2}}\right)^{2}.\label{eq: w_xy_ax}
\end{equation}
Subsequently, by substituting Eq.(\ref{eq: A_ax}) into Eq.(\ref{eq: Friedmann_ax_xy}),
we derive the scale factor $C\left(t\right)$ as follows;

\begin{equation}
C\left(t\right)=\frac{C_{3}\left(C_{1}e^{\sqrt{3\tilde{\Lambda}}t}-C_{2}\right)^{\frac{5}{6}}\left(C_{1}^{2}e^{2\sqrt{3\tilde{\Lambda}}t}-C_{2}^{2}\right)^{\frac{1}{6}}}{e^{\frac{\sqrt{3\tilde{\Lambda}}t}{3}}\sqrt{C_{1}e^{\sqrt{3\tilde{\Lambda}}t}+C_{2}}},\label{eq: C_ax}
\end{equation}
Using this expression, we can determine the Hubble parameter $H_{z}\left(t\right)$
and the effective equation of state parameter $w_{z}\left(t\right)$
for the $z$-direction:
\begin{equation}
H_{z}\left(t\right)=-\frac{4C_{1}C_{2}\tilde{\Lambda}e^{\sqrt{3\tilde{\Lambda}}t}\left(C_{1}^{2}e^{2\sqrt{3\tilde{\Lambda}}t}+C_{1}C_{2}e^{\sqrt{3\tilde{\Lambda}}t}+C_{2}^{2}\right)}{\left(C_{1}^{2}e^{2\sqrt{3\tilde{\Lambda}}t}-C_{2}^{2}\right)^{2}},\label{eq: H_ax_z}
\end{equation}

\begin{equation}
w_{z}\left(t\right)=-\frac{C_{1}^{4}e^{4\sqrt{3\tilde{\Lambda}}t}+10C_{1}^{2}C_{2}^{2}e^{2\sqrt{3\tilde{\Lambda}}t}+C_{2}^{4}}{\left(C_{1}^{2}e^{2\sqrt{3\tilde{\Lambda}}t}+4C_{1}C_{2}e^{\sqrt{3\tilde{\Lambda}}t}+C_{2}^{2}\right)^{2}}.\label{eq: w_ax_z}
\end{equation}

\begin{figure}
\includegraphics[width=9cm]{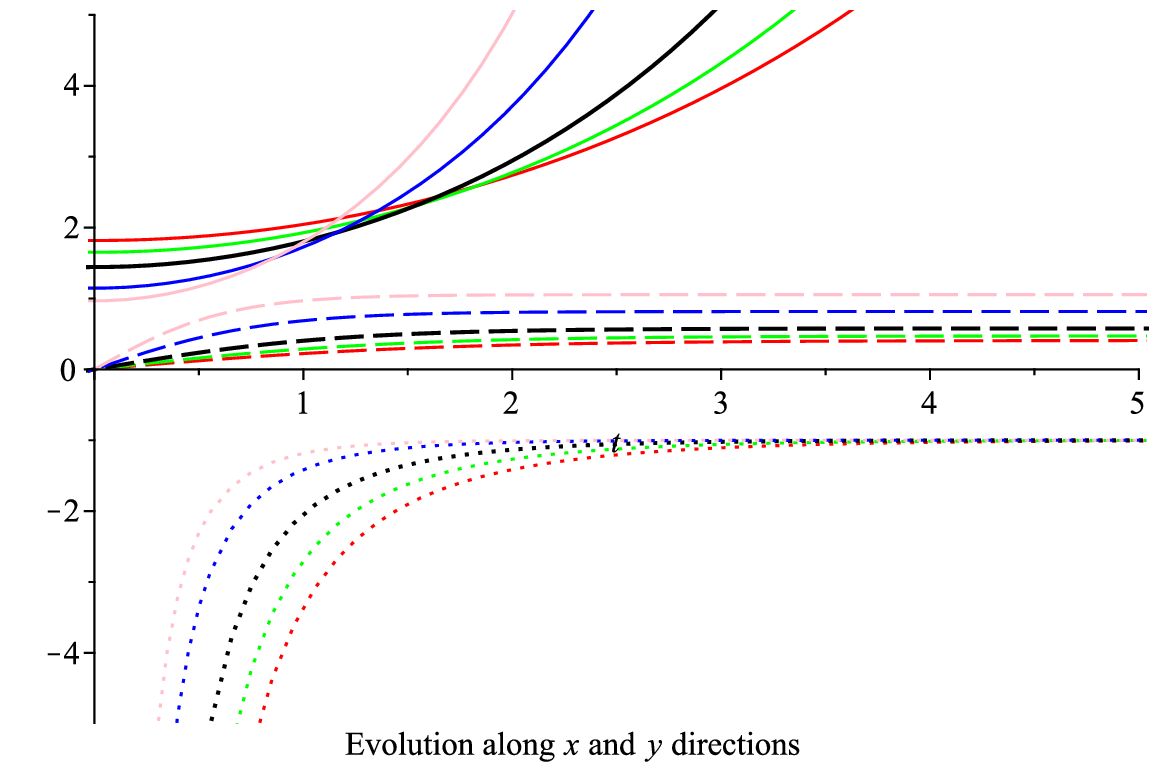}\includegraphics[width=9cm]{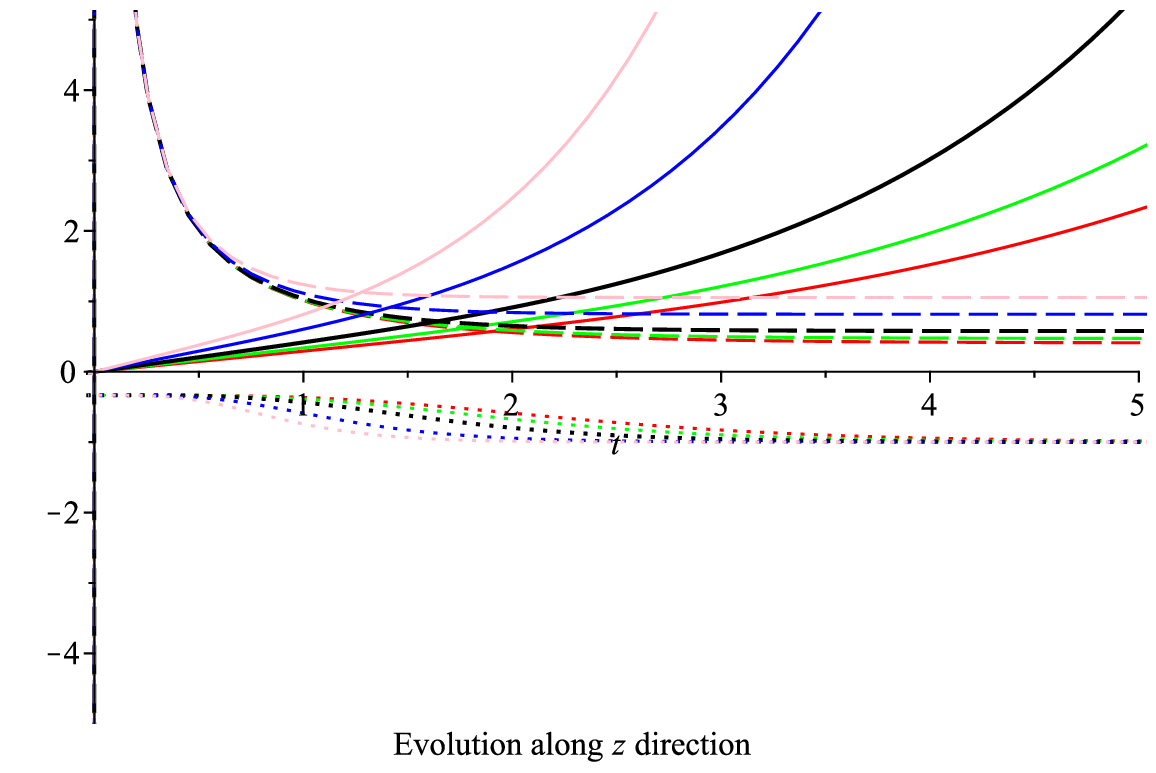}

\caption{(color online) These plots show the evolution of the cosmological
parameters $A\left(t\right)$ (solid line), $H_{xy}\left(t\right)$
(dashed line), $w_{xy}\left(t\right)$ (dotted line) along the $x$
and $y$ directions and $C\left(t\right)$ (solid line), $H_{z}\left(t\right)$
(dashed line), $w_{z}\left(t\right)$ (dotted line) along the $z$
direction. In these graphs we compare the evolution of the given parameters
with different values of $\alpha$ deformation parameter ($\alpha=1$
(black) is the non deformed case, $\alpha=2$ (red), $\alpha=1.5$
(green), $\alpha=0.5$ (blue) and $\alpha=0.3$ (pink) for deformed
cases) under the constraints $\Lambda=1$, $C_{1}=1$, $C_{2}=1$
and $C_{3}=0.3$.\label{fig:General_Plot}}
\end{figure}

\begin{figure}
\includegraphics[width=9cm]{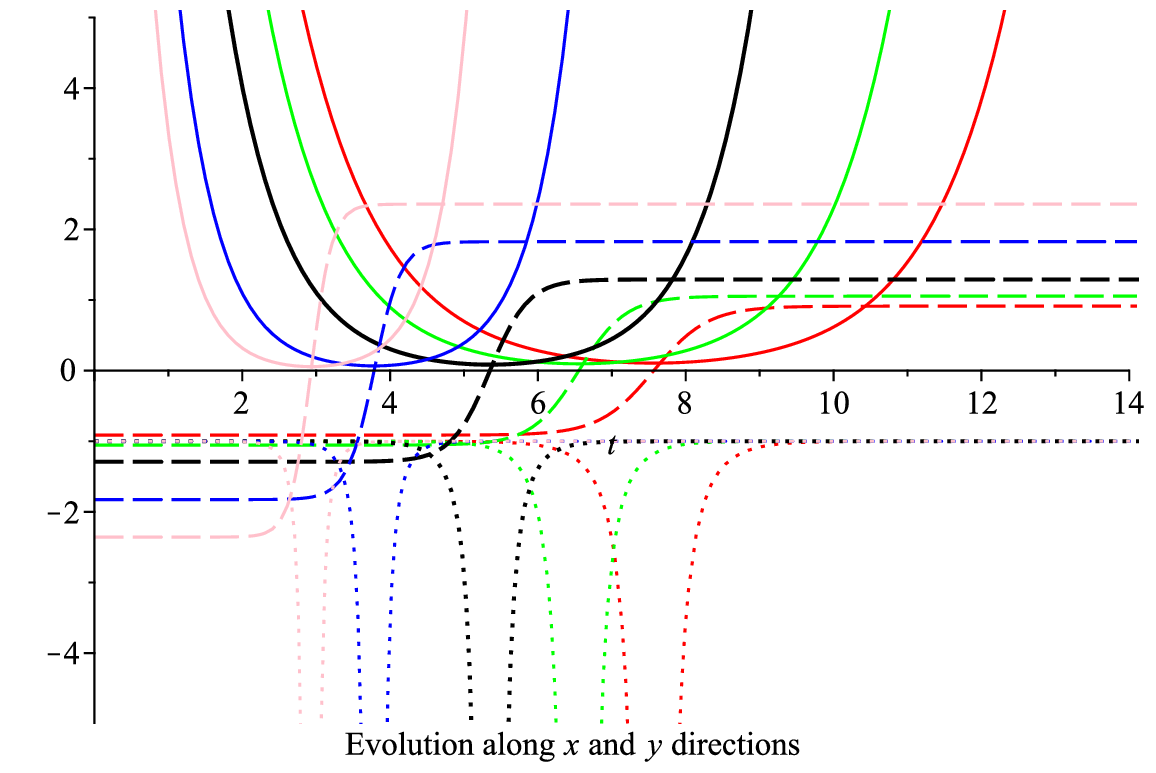}\includegraphics[width=9cm]{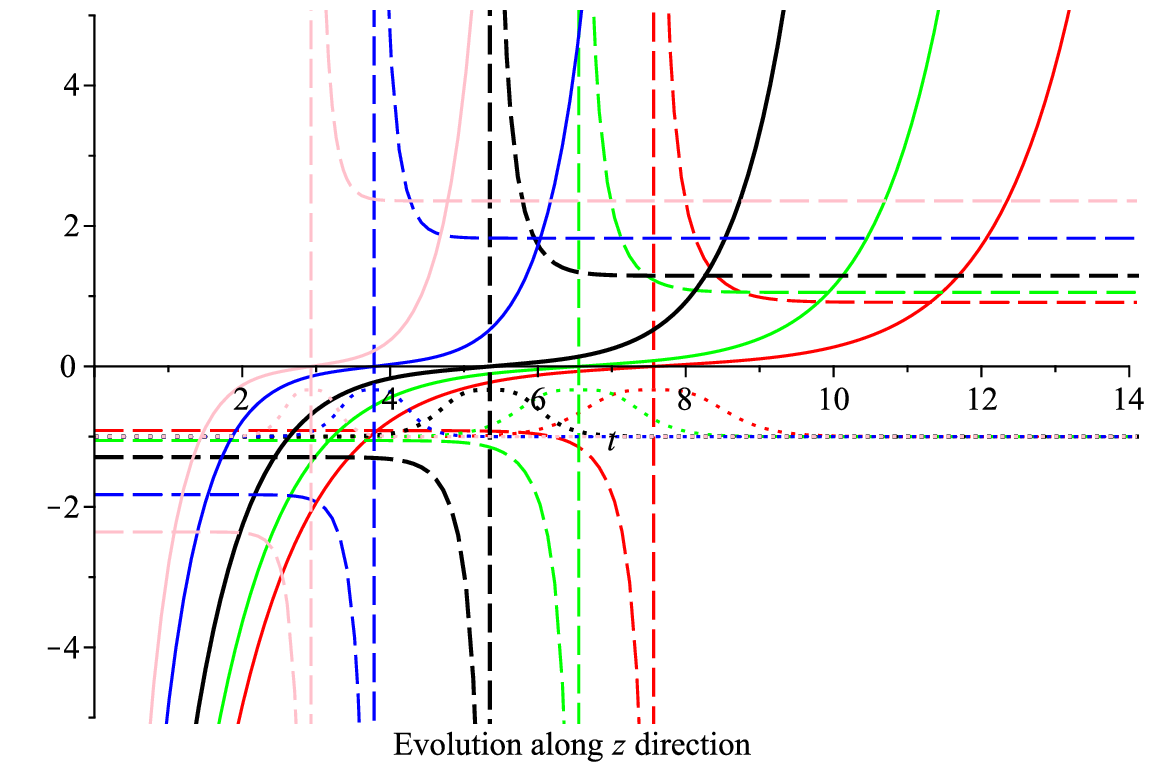}

\caption{(color online) These plots show the evolution of the cosmological
parameters $A\left(t\right)$ (solid line), $H_{xy}\left(t\right)$
(dashed line), $w_{xy}\left(t\right)$ (dotted line) along the $x$
and $y$ directions and $C\left(t\right)$ (solid line), $H_{z}\left(t\right)$
(dashed line), $w_{z}\left(t\right)$ (dotted line) along the $z$
direction. In these graphs we compare the evolution of the given parameters
with different values of $\alpha$ deformation parameter ($\alpha=1$
(black) is the non deformed case, $\alpha=2$ (red), $\alpha=1.5$
(green), $\alpha=0.5$ (blue) and $\alpha=0.3$ (pink) for deformed
cases) under the constraints $\Lambda=5$, $C_{1}=10^{-6}$, $C_{2}=10^{3}$
and $C_{3}=0.3$.\label{fig: Bounce_Plot}}
\end{figure}

Finally, we illustrate the evolution of key cosmological parameters,
including the scale factor, the Hubble parameter, and the effective
equation of state parameter, under different limiting conditions in
Figures \ref{fig:General_Plot} and \ref{fig: Bounce_Plot}. These
figures offer a comparative analysis between the deformed and standard
cosmological evolutions ($\alpha\left(z,q\right)=1$) of these parameters. 

\section{Conclusion\label{sec:Conclusion}}

In this paper, we have explored the implications of $q$-deformed
fermions within the context of anisotropic cosmology models, utilizing
Verlinde's entropic gravity as the foundational framework. Our investigation
reveals that the incorporation of $q$-deformations, a concept rooted
in quantum groups, introduces notable deviations from classical cosmological
predictions, particularly in the dynamics of the early universe and
its subsequent evolution.

Our analysis demonstrates that such deformations exert a significant
influence on the dynamics of anisotropic cosmologies, resulting in
modified equations governing the evolution of the universe, as shown
in Eqs.(\ref{eq: Friedmann_t})-(\ref{eq: Friedmann_acc}). Focusing
on the axisymmetric solution, we derived $q$-deformed versions of
the Friedmann equations, presented in Eqs.(\ref{eq: Friedmann_ax_t})-(\ref{eq: Friedmann_ax_z})
which govern the evolution of the anisotropic scale factors. Furthermore,
we obtained exact analytical solutions for the scale parameters, Hubble
parameters, and the effective equation of state parameters, for both
the $x$- and $y$-directions in Eqs.(\ref{eq: A_ax})-(\ref{eq: w_xy_ax})
and and along the $z$-direction in Eqs.(\ref{eq: C_ax})-(\ref{eq: w_ax_z}).

As shown in Figure \ref{fig:General_Plot}, for larger values of $\alpha\left(z,q\right)$,
the system evolves more quickly, with faster expansion rates for the
scale factors and quicker stabilization of the Hubble parameters.
Both the EoS parameters in the $x$, $y$, and $z$ directions show
that the system tends towards a vacuum-dominated phase ($w\approx-1$),
but deformation affects how quickly this occurs. Anisotropy is evident
in how the dynamics differ between the $x$-$y$ directions and the
$z$ direction, especially in the early universe, where different
values of $\alpha\left(z,q\right)$ lead to varying initial expansion
behaviors. 

Figure \ref{fig: Bounce_Plot} highlights that at the higher values
of $\alpha\left(z,q\right)$, there are sharp fluctuations in both
the Hubble parameters and the equation of state, indicating that the
deformation introduces more complex evolution patterns. However, for
lower deformation values ($\alpha\left(z,q\right)=0.3$ and $\alpha\left(z,q\right)=0.5$),
the universe's behavior stabilizes more quickly. This graph suggests
that $q$-deformation, depending on its strength, can introduce significant
anisotropy in the early stages of cosmological evolution, with long-term
behavior converging to a vacuum-like state. However, strong deformation
may lead to transient phases of accelerated expansion, especially
in the $z$ direction. Furthermore the evolution of $A\left(t\right)$
in the graph demonstrates the hallmarks of a non-singular bounce cosmology
\citep{Peter:2002Primordial,Allen:2004Cosmological,Cai:2011MatterBounce,Cai:2012Towards,Nojiri:2019NonsingularBounce,Odintsov:2020FromBounce,Kumar:2021Anisotropic}.
The initial decrease corresponds to a contraction phase, the minimum
point of $A\left(t\right)$ marks the bounce itself, and the subsequent
increase reflects the expansion phase of the universe. The role of
the deformation parameter $\alpha\left(z,q\right)$ further modifies
the nature of this bounce, influencing both the timing and the dynamics
of the transition between contraction and expansion. This behavior
is consistent with many bounce cosmology models, where modifications
to gravity or quantum effects prevent singularity and allow the universe
to smoothly transition through a bounce.

In summary, our results demonstrate that the inclusion of $q$-deformations
introduces significant deviations from classical cosmological predictions,
particularly in the early universe's dynamics. These deformations
offer a viable framework for addressing key cosmological challenges,
including singularity avoidance and the explanation of early anisotropies
in the universe\textquoteright s expansion. The findings suggest that
$q$-deformation may play a critical role in understanding the universe's
initial conditions and its evolution, offering new avenues for exploring
the effects of quantum modifications in cosmology. Furthermore, it
would be of great interest to investigate the unification of various
cosmological epochs within the framework of $q$-deformed gravity
under anisotropic conditions, a topic we intend to explore in future
studies. It is also noteworthy that the application of $q$-deformation
has not been extensively explored in the context of the Hubble tension,
$H_{0}$ \citep{Bernal:2016TheTroubleH0}. The recent review by\citep{Vagnozzi:2023Seven}
examines modifications to early and late universe models as potential
solutions to the current cosmic tensions, a subject that will be addressed
in future research. On the other hand, an alternative approach to
Verlinde's entropic gravity approximation involves modeling cosmological
phenomena through using a generalized entropy function \citep{Nojiri:2022Early,Nojiri:2022Modified,Nojiri:2024Different}.
This framework integrates diverse entropy formulations, including
Tsallis, Rényi, Barrow, Kaniadakis, Sharma-Mittal, and those arising
from loop quantum gravity, into a cohesive structure. Such a model
suggests a connection between early inflation and the late dark energy
era, and investigating the implications of this model in anisotropic
cosmology might be an interesting subject for future studies. We aim
to tackle some of the aforementioned issues in subsequent work.

\section*{Declaration of competing interest }

The authors declare that they have no known competing financial interests
or personal relationships that could have appeared to influence the
work reported in this paper.

\section*{Data availability }

No data was used for the research described in the article.

\section*{Declaration of generative AI and AI-assisted technologies in the
writing process}

During the preparation of this work the author used ChatGPT - GPT-4o
in order to improve readability of the text. After using this tool/service,
the author reviewed and edited the content as needed and takes full
responsibility for the content of the publication.

\bibliography{q_def_anisotropic}

\begin{thebibliography}{71}%
\makeatletter
\providecommand \@ifxundefined [1]{%
 \@ifx{#1\undefined}
}%
\providecommand \@ifnum [1]{%
 \ifnum #1\expandafter \@firstoftwo
 \else \expandafter \@secondoftwo
 \fi
}%
\providecommand \@ifx [1]{%
 \ifx #1\expandafter \@firstoftwo
 \else \expandafter \@secondoftwo
 \fi
}%
\providecommand \natexlab [1]{#1}%
\providecommand \enquote  [1]{``#1''}%
\providecommand \bibnamefont  [1]{#1}%
\providecommand \bibfnamefont [1]{#1}%
\providecommand \citenamefont [1]{#1}%
\providecommand \href@noop [0]{\@secondoftwo}%
\providecommand \href [0]{\begingroup \@sanitize@url \@href}%
\providecommand \@href[1]{\@@startlink{#1}\@@href}%
\providecommand \@@href[1]{\endgroup#1\@@endlink}%
\providecommand \@sanitize@url [0]{\catcode `\\12\catcode `\$12\catcode
  `\&12\catcode `\#12\catcode `\^12\catcode `\_12\catcode `\%12\relax}%
\providecommand \@@startlink[1]{}%
\providecommand \@@endlink[0]{}%
\providecommand \url  [0]{\begingroup\@sanitize@url \@url }%
\providecommand \@url [1]{\endgroup\@href {#1}{\urlprefix }}%
\providecommand \urlprefix  [0]{URL }%
\providecommand \Eprint [0]{\href }%
\providecommand \doibase [0]{https://doi.org/}%
\providecommand \selectlanguage [0]{\@gobble}%
\providecommand \bibinfo  [0]{\@secondoftwo}%
\providecommand \bibfield  [0]{\@secondoftwo}%
\providecommand \translation [1]{[#1]}%
\providecommand \BibitemOpen [0]{}%
\providecommand \bibitemStop [0]{}%
\providecommand \bibitemNoStop [0]{.\EOS\space}%
\providecommand \EOS [0]{\spacefactor3000\relax}%
\providecommand \BibitemShut  [1]{\csname bibitem#1\endcsname}%
\let\auto@bib@innerbib\@empty
\bibitem [{\citenamefont {Capozziello}\ and\ \citenamefont
  {De~Laurentis}(2011)}]{Capozziello:2011Extended}%
  \BibitemOpen
  \bibfield  {author} {\bibinfo {author} {\bibfnamefont {S.}~\bibnamefont
  {Capozziello}}\ and\ \bibinfo {author} {\bibfnamefont {M.}~\bibnamefont
  {De~Laurentis}},\ }\bibfield  {title} {\bibinfo {title} {{Extended Theories
  of Gravity}},\ }\href {https://doi.org/10.1016/j.physrep.2011.09.003}
  {\bibfield  {journal} {\bibinfo  {journal} {Phys. Rept.}\ }\textbf {\bibinfo
  {volume} {509}},\ \bibinfo {pages} {167} (\bibinfo {year}
  {2011})}\BibitemShut {NoStop}%
\bibitem [{\citenamefont {Clifton}\ \emph {et~al.}(2012)\citenamefont
  {Clifton}, \citenamefont {Ferreira}, \citenamefont {Padilla},\ and\
  \citenamefont {Skordis}}]{Clifton:2011Modified}%
  \BibitemOpen
  \bibfield  {author} {\bibinfo {author} {\bibfnamefont {T.}~\bibnamefont
  {Clifton}}, \bibinfo {author} {\bibfnamefont {P.~G.}\ \bibnamefont
  {Ferreira}}, \bibinfo {author} {\bibfnamefont {A.}~\bibnamefont {Padilla}},\
  and\ \bibinfo {author} {\bibfnamefont {C.}~\bibnamefont {Skordis}},\
  }\bibfield  {title} {\bibinfo {title} {{Modified Gravity and Cosmology}},\
  }\href {https://doi.org/10.1016/j.physrep.2012.01.001} {\bibfield  {journal}
  {\bibinfo  {journal} {Phys. Rept.}\ }\textbf {\bibinfo {volume} {513}},\
  \bibinfo {pages} {1} (\bibinfo {year} {2012})}\BibitemShut {NoStop}%
\bibitem [{\citenamefont {Nojiri}\ \emph {et~al.}(2017)\citenamefont {Nojiri},
  \citenamefont {Odintsov},\ and\ \citenamefont
  {Oikonomou}}]{Nojiri:2017Modified}%
  \BibitemOpen
  \bibfield  {author} {\bibinfo {author} {\bibfnamefont {S.}~\bibnamefont
  {Nojiri}}, \bibinfo {author} {\bibfnamefont {S.~D.}\ \bibnamefont
  {Odintsov}},\ and\ \bibinfo {author} {\bibfnamefont {V.~K.}\ \bibnamefont
  {Oikonomou}},\ }\bibfield  {title} {\bibinfo {title} {{Modified Gravity
  Theories on a Nutshell: Inflation, Bounce and Late-time Evolution}},\ }\href
  {https://doi.org/10.1016/j.physrep.2017.06.001} {\bibfield  {journal}
  {\bibinfo  {journal} {Phys. Rept.}\ }\textbf {\bibinfo {volume} {692}},\
  \bibinfo {pages} {1} (\bibinfo {year} {2017})}\BibitemShut {NoStop}%
\bibitem [{\citenamefont {Arik}\ and\ \citenamefont
  {Coon}(1976)}]{Arik:1976Hilbert}%
  \BibitemOpen
  \bibfield  {author} {\bibinfo {author} {\bibfnamefont {M.}~\bibnamefont
  {Arik}}\ and\ \bibinfo {author} {\bibfnamefont {D.~D.}\ \bibnamefont
  {Coon}},\ }\bibfield  {title} {\bibinfo {title} {{Hilbert Spaces of Analytic
  Functions and Generalized Coherent States}},\ }\href
  {https://doi.org/10.1063/1.522937} {\bibfield  {journal} {\bibinfo  {journal}
  {J. Math. Phys.}\ }\textbf {\bibinfo {volume} {17}},\ \bibinfo {pages} {524}
  (\bibinfo {year} {1976})}\BibitemShut {NoStop}%
\bibitem [{\citenamefont {Tsallis}(1988)}]{Tsallis:1988Possible}%
  \BibitemOpen
  \bibfield  {author} {\bibinfo {author} {\bibfnamefont {C.}~\bibnamefont
  {Tsallis}},\ }\bibfield  {title} {\bibinfo {title} {{Possible Generalization
  of Boltzmann-Gibbs Statistics}},\ }\href {https://doi.org/10.1007/BF01016429}
  {\bibfield  {journal} {\bibinfo  {journal} {J. Statist. Phys.}\ }\textbf
  {\bibinfo {volume} {52}},\ \bibinfo {pages} {479} (\bibinfo {year}
  {1988})}\BibitemShut {NoStop}%
\bibitem [{\citenamefont {Biedenharn}(1989)}]{Biedenharn:1989TheQuantum}%
  \BibitemOpen
  \bibfield  {author} {\bibinfo {author} {\bibfnamefont {L.~C.}\ \bibnamefont
  {Biedenharn}},\ }\bibfield  {title} {\bibinfo {title} {{The Quantum Group
  SU(2)-q and a q Analog of the Boson Operators}},\ }\href
  {https://doi.org/10.1088/0305-4470/22/18/004} {\bibfield  {journal} {\bibinfo
   {journal} {J. Phys. A}\ }\textbf {\bibinfo {volume} {22}},\ \bibinfo {pages}
  {L873} (\bibinfo {year} {1989})}\BibitemShut {NoStop}%
\bibitem [{\citenamefont {Macfarlane}(1989)}]{Macfarlane:1989OnQAnalogs}%
  \BibitemOpen
  \bibfield  {author} {\bibinfo {author} {\bibfnamefont {A.~J.}\ \bibnamefont
  {Macfarlane}},\ }\bibfield  {title} {\bibinfo {title} {{On q Analogs of the
  Quantum Harmonic Oscillator and the Quantum Group SU(2)-q}},\ }\href
  {https://doi.org/10.1088/0305-4470/22/21/020} {\bibfield  {journal} {\bibinfo
   {journal} {J. Phys. A}\ }\textbf {\bibinfo {volume} {22}},\ \bibinfo {pages}
  {4581} (\bibinfo {year} {1989})}\BibitemShut {NoStop}%
\bibitem [{\citenamefont {Sun}\ and\ \citenamefont
  {Fu}(1989)}]{Sun:1989TheQDeformed}%
  \BibitemOpen
  \bibfield  {author} {\bibinfo {author} {\bibfnamefont {C.-P.}\ \bibnamefont
  {Sun}}\ and\ \bibinfo {author} {\bibfnamefont {H.-C.}\ \bibnamefont {Fu}},\
  }\bibfield  {title} {\bibinfo {title} {{The q deformed boson realization of
  the quantum group SU(n)-q and its representation}},\ }\href
  {https://doi.org/10.1088/0305-4470/22/21/001} {\bibfield  {journal} {\bibinfo
   {journal} {J. Phys. A}\ }\textbf {\bibinfo {volume} {22}},\ \bibinfo {pages}
  {L983} (\bibinfo {year} {1989})}\BibitemShut {NoStop}%
\bibitem [{\citenamefont {Lavagno}\ and\ \citenamefont
  {Narayana~Swamy}(2002)}]{Lavagno:2002Generalized}%
  \BibitemOpen
  \bibfield  {author} {\bibinfo {author} {\bibfnamefont {A.}~\bibnamefont
  {Lavagno}}\ and\ \bibinfo {author} {\bibfnamefont {P.}~\bibnamefont
  {Narayana~Swamy}},\ }\bibfield  {title} {\bibinfo {title} {{Generalized
  thermodynamics of q deformed bosons and fermions}},\ }\href
  {https://doi.org/10.1103/PhysRevE.65.036101} {\bibfield  {journal} {\bibinfo
  {journal} {Phys. Rev. E}\ }\textbf {\bibinfo {volume} {65}},\ \bibinfo
  {pages} {036101} (\bibinfo {year} {2002})}\BibitemShut {NoStop}%
\bibitem [{\citenamefont {Jalalzadeh}\ \emph {et~al.}(2023)\citenamefont
  {Jalalzadeh}, \citenamefont {Moradpour},\ and\ \citenamefont
  {Moniz}}]{Jalalzadeh:2023Modified}%
  \BibitemOpen
  \bibfield  {author} {\bibinfo {author} {\bibfnamefont {S.}~\bibnamefont
  {Jalalzadeh}}, \bibinfo {author} {\bibfnamefont {H.}~\bibnamefont
  {Moradpour}},\ and\ \bibinfo {author} {\bibfnamefont {P.~V.}\ \bibnamefont
  {Moniz}},\ }\bibfield  {title} {\bibinfo {title} {{Modified cosmology from
  quantum deformed entropy}},\ }\href
  {https://doi.org/10.1016/j.dark.2023.101320} {\bibfield  {journal} {\bibinfo
  {journal} {Phys. Dark Univ.}\ }\textbf {\bibinfo {volume} {42}},\ \bibinfo
  {pages} {101320} (\bibinfo {year} {2023})}\BibitemShut {NoStop}%
\bibitem [{\citenamefont {Nojiri}\ \emph {et~al.}(2023)\citenamefont {Nojiri},
  \citenamefont {Odintsov},\ and\ \citenamefont
  {Paul}}]{Nojiri:2023Microscopic}%
  \BibitemOpen
  \bibfield  {author} {\bibinfo {author} {\bibfnamefont {S.}~\bibnamefont
  {Nojiri}}, \bibinfo {author} {\bibfnamefont {S.~D.}\ \bibnamefont
  {Odintsov}},\ and\ \bibinfo {author} {\bibfnamefont {T.}~\bibnamefont
  {Paul}},\ }\bibfield  {title} {\bibinfo {title} {{Microscopic interpretation
  of generalized entropy}},\ }\href
  {https://doi.org/10.1016/j.physletb.2023.138321} {\bibfield  {journal}
  {\bibinfo  {journal} {Phys. Lett. B}\ }\textbf {\bibinfo {volume} {847}},\
  \bibinfo {pages} {138321} (\bibinfo {year} {2023})}\BibitemShut {NoStop}%
\bibitem [{\citenamefont {Wilczek}(1990)}]{Wilczek:1990Fractional}%
  \BibitemOpen
  \bibinfo {editor} {\bibfnamefont {F.}~\bibnamefont {Wilczek}},\ ed.,\
  \href@noop {} {\emph {\bibinfo {title} {{Fractional statistics and anyon
  superconductivity}}}}\ (\bibinfo {year} {1990})\BibitemShut {NoStop}%
\bibitem [{\citenamefont {Brito}\ and\ \citenamefont
  {Marinho}(2011)}]{Brito:2011q-Deformed}%
  \BibitemOpen
  \bibfield  {author} {\bibinfo {author} {\bibfnamefont {F.~A.}\ \bibnamefont
  {Brito}}\ and\ \bibinfo {author} {\bibfnamefont {A.~A.}\ \bibnamefont
  {Marinho}},\ }\bibfield  {title} {\bibinfo {title} {{q-Deformed Landau
  diamagnetism problem embedded in D-dimensions}},\ }\href
  {https://doi.org/10.1016/j.physa.2011.03.003} {\bibfield  {journal} {\bibinfo
   {journal} {Physica A}\ }\textbf {\bibinfo {volume} {390}},\ \bibinfo {pages}
  {2497} (\bibinfo {year} {2011})}\BibitemShut {NoStop}%
\bibitem [{\citenamefont {Bonatsos}\ and\ \citenamefont
  {Daskaloyannis}(1999)}]{Bonatsos:1999Quantum}%
  \BibitemOpen
  \bibfield  {author} {\bibinfo {author} {\bibfnamefont {D.}~\bibnamefont
  {Bonatsos}}\ and\ \bibinfo {author} {\bibfnamefont {C.}~\bibnamefont
  {Daskaloyannis}},\ }\bibfield  {title} {\bibinfo {title} {{Quantum groups and
  their applications in nuclear physics}},\ }\href
  {https://doi.org/10.1016/S0146-6410(99)00100-3} {\bibfield  {journal}
  {\bibinfo  {journal} {Prog. Part. Nucl. Phys.}\ }\textbf {\bibinfo {volume}
  {43}},\ \bibinfo {pages} {537} (\bibinfo {year} {1999})}\BibitemShut
  {NoStop}%
\bibitem [{\citenamefont {\c{S}enay}\ and\ \citenamefont
  {Kibaro\u{g}lu}(2019)}]{Senay:2018xaj}%
  \BibitemOpen
  \bibfield  {author} {\bibinfo {author} {\bibfnamefont {M.}~\bibnamefont
  {\c{S}enay}}\ and\ \bibinfo {author} {\bibfnamefont {S.}~\bibnamefont
  {Kibaro\u{g}lu}},\ }\bibfield  {title} {\bibinfo {title} {{$q$-Deformed
  Einstein equations from entropic force}},\ }\href
  {https://doi.org/10.1142/S0217751X18502184} {\bibfield  {journal} {\bibinfo
  {journal} {Int. J. Mod. Phys. A}\ }\textbf {\bibinfo {volume} {33}},\
  \bibinfo {pages} {1850218} (\bibinfo {year} {2019})}\BibitemShut {NoStop}%
\bibitem [{\citenamefont {Kibaro\u{g}lu}\ and\ \citenamefont
  {Senay}(2019)}]{Kibaroglu:2018mnx}%
  \BibitemOpen
  \bibfield  {author} {\bibinfo {author} {\bibfnamefont {S.}~\bibnamefont
  {Kibaro\u{g}lu}}\ and\ \bibinfo {author} {\bibfnamefont {M.}~\bibnamefont
  {Senay}},\ }\bibfield  {title} {\bibinfo {title} {{Effects of bosonic and
  fermionic $q$-deformation on the entropic gravity}},\ }\href
  {https://doi.org/10.1142/S0217732319502493} {\bibfield  {journal} {\bibinfo
  {journal} {Mod. Phys. Lett. A}\ }\textbf {\bibinfo {volume} {34}},\ \bibinfo
  {pages} {1950249} (\bibinfo {year} {2019})}\BibitemShut {NoStop}%
\bibitem [{\citenamefont {Kibaro\u{g}lu}\ and\ \citenamefont
  {Senay}(2020)}]{Kibaroglu:2019odt}%
  \BibitemOpen
  \bibfield  {author} {\bibinfo {author} {\bibfnamefont {S.}~\bibnamefont
  {Kibaro\u{g}lu}}\ and\ \bibinfo {author} {\bibfnamefont {M.}~\bibnamefont
  {Senay}},\ }\bibfield  {title} {\bibinfo {title} {{Friedmann equations for
  deformed entropic gravity}},\ }\href
  {https://doi.org/10.1142/S021827182050042X} {\bibfield  {journal} {\bibinfo
  {journal} {Int. J. Mod. Phys. D}\ }\textbf {\bibinfo {volume} {29}},\
  \bibinfo {pages} {2050042} (\bibinfo {year} {2020})}\BibitemShut {NoStop}%
\bibitem [{\citenamefont {Senay}(2024{\natexlab{a}})}]{Senay:2024Jeans}%
  \BibitemOpen
  \bibfield  {author} {\bibinfo {author} {\bibfnamefont {M.}~\bibnamefont
  {Senay}},\ }\bibfield  {title} {\bibinfo {title} {{Jeans mass and Gamow
  temperature: insights from q-Deformed systems}},\ }\href
  {https://doi.org/10.1088/1402-4896/ad7176} {\bibfield  {journal} {\bibinfo
  {journal} {Phys. Scripta}\ }\textbf {\bibinfo {volume} {99}},\ \bibinfo
  {pages} {105001} (\bibinfo {year} {2024}{\natexlab{a}})}\BibitemShut
  {NoStop}%
\bibitem [{\citenamefont {Senay}(2021)}]{Senay:2021Entropic}%
  \BibitemOpen
  \bibfield  {author} {\bibinfo {author} {\bibfnamefont {M.}~\bibnamefont
  {Senay}},\ }\bibfield  {title} {\bibinfo {title} {{Entropic gravity corrected
  by q-statistics, and its implications to cosmology}},\ }\href
  {https://doi.org/10.1016/j.physletb.2021.136536} {\bibfield  {journal}
  {\bibinfo  {journal} {Phys. Lett. B}\ }\textbf {\bibinfo {volume} {820}},\
  \bibinfo {pages} {136536} (\bibinfo {year} {2021})}\BibitemShut {NoStop}%
\bibitem [{\citenamefont {Senay}\ \emph {et~al.}(2021)\citenamefont {Senay},
  \citenamefont {Mohammadi~Sabet},\ and\ \citenamefont
  {Moradpour}}]{Senay:2021Heat}%
  \BibitemOpen
  \bibfield  {author} {\bibinfo {author} {\bibfnamefont {M.}~\bibnamefont
  {Senay}}, \bibinfo {author} {\bibfnamefont {M.}~\bibnamefont
  {Mohammadi~Sabet}},\ and\ \bibinfo {author} {\bibfnamefont {H.}~\bibnamefont
  {Moradpour}},\ }\bibfield  {title} {\bibinfo {title} {{Heat capacity of
  Holographic screen inspires MOND theory}},\ }\href
  {https://doi.org/10.1088/1402-4896/abf618} {\bibfield  {journal} {\bibinfo
  {journal} {Phys. Scripta}\ }\textbf {\bibinfo {volume} {96}},\ \bibinfo
  {pages} {075001} (\bibinfo {year} {2021})}\BibitemShut {NoStop}%
\bibitem [{\citenamefont {Senay}(2024{\natexlab{b}})}]{Senay:2024TheRoleMond}%
  \BibitemOpen
  \bibfield  {author} {\bibinfo {author} {\bibfnamefont {M.}~\bibnamefont
  {Senay}},\ }\bibfield  {title} {\bibinfo {title} {{The Role of MOND Theory in
  Cosmological Paradigms Under q-Statistics}},\ }\href
  {https://doi.org/10.1007/s10773-024-05801-w} {\bibfield  {journal} {\bibinfo
  {journal} {Int. J. Theor. Phys.}\ }\textbf {\bibinfo {volume} {63}},\
  \bibinfo {pages} {263} (\bibinfo {year} {2024}{\natexlab{b}})}\BibitemShut
  {NoStop}%
\bibitem [{\citenamefont {Durrer}(2015)}]{Durrer:2015TheCosmic}%
  \BibitemOpen
  \bibfield  {author} {\bibinfo {author} {\bibfnamefont {R.}~\bibnamefont
  {Durrer}},\ }\bibfield  {title} {\bibinfo {title} {{The cosmic microwave
  background: the history of its experimental investigation and its
  significance for cosmology}},\ }\href
  {https://doi.org/10.1088/0264-9381/32/12/124007} {\bibfield  {journal}
  {\bibinfo  {journal} {Class. Quant. Grav.}\ }\textbf {\bibinfo {volume}
  {32}},\ \bibinfo {pages} {124007} (\bibinfo {year} {2015})}\BibitemShut
  {NoStop}%
\bibitem [{\citenamefont {Vakili}\ and\ \citenamefont
  {Khosravi}(2012)}]{Vakili:2012Classical}%
  \BibitemOpen
  \bibfield  {author} {\bibinfo {author} {\bibfnamefont {B.}~\bibnamefont
  {Vakili}}\ and\ \bibinfo {author} {\bibfnamefont {N.}~\bibnamefont
  {Khosravi}},\ }\bibfield  {title} {\bibinfo {title} {{Classical and quantum
  massive cosmology for the open FRW universe}},\ }\href
  {https://doi.org/10.1103/PhysRevD.85.083529} {\bibfield  {journal} {\bibinfo
  {journal} {Phys. Rev. D}\ }\textbf {\bibinfo {volume} {85}},\ \bibinfo
  {pages} {083529} (\bibinfo {year} {2012})}\BibitemShut {NoStop}%
\bibitem [{\citenamefont {Hinshaw}\ \emph {et~al.}(2003)\citenamefont {Hinshaw}
  \emph {et~al.}}]{WMAP:2003}%
  \BibitemOpen
  \bibfield  {author} {\bibinfo {author} {\bibfnamefont {G.}~\bibnamefont
  {Hinshaw}} \emph {et~al.} (\bibinfo {collaboration} {WMAP}),\ }\bibfield
  {title} {\bibinfo {title} {{First year Wilkinson Microwave Anisotropy Probe
  (WMAP) observations: The Angular power spectrum}},\ }\href
  {https://doi.org/10.1086/377225} {\bibfield  {journal} {\bibinfo  {journal}
  {Astrophys. J. Suppl.}\ }\textbf {\bibinfo {volume} {148}},\ \bibinfo {pages}
  {135} (\bibinfo {year} {2003})}\BibitemShut {NoStop}%
\bibitem [{\citenamefont {Hinshaw}\ \emph {et~al.}(2007)\citenamefont {Hinshaw}
  \emph {et~al.}}]{WMAP:2006}%
  \BibitemOpen
  \bibfield  {author} {\bibinfo {author} {\bibfnamefont {G.}~\bibnamefont
  {Hinshaw}} \emph {et~al.} (\bibinfo {collaboration} {WMAP}),\ }\bibfield
  {title} {\bibinfo {title} {{Three-year Wilkinson Microwave Anisotropy Probe
  (WMAP) observations: temperature analysis}},\ }\href
  {https://doi.org/10.1086/513698} {\bibfield  {journal} {\bibinfo  {journal}
  {Astrophys. J. Suppl.}\ }\textbf {\bibinfo {volume} {170}},\ \bibinfo {pages}
  {288} (\bibinfo {year} {2007})}\BibitemShut {NoStop}%
\bibitem [{\citenamefont {Hinshaw}\ \emph {et~al.}(2009)\citenamefont {Hinshaw}
  \emph {et~al.}}]{WMAP:2008}%
  \BibitemOpen
  \bibfield  {author} {\bibinfo {author} {\bibfnamefont {G.}~\bibnamefont
  {Hinshaw}} \emph {et~al.} (\bibinfo {collaboration} {WMAP}),\ }\bibfield
  {title} {\bibinfo {title} {{Five-Year Wilkinson Microwave Anisotropy Probe
  (WMAP) Observations: Data Processing, Sky Maps, and Basic Results}},\ }\href
  {https://doi.org/10.1088/0067-0049/180/2/225} {\bibfield  {journal} {\bibinfo
   {journal} {Astrophys. J. Suppl.}\ }\textbf {\bibinfo {volume} {180}},\
  \bibinfo {pages} {225} (\bibinfo {year} {2009})}\BibitemShut {NoStop}%
\bibitem [{\citenamefont {Jaffe}\ \emph {et~al.}(2005)\citenamefont {Jaffe},
  \citenamefont {Banday}, \citenamefont {Eriksen}, \citenamefont {Gorski},\
  and\ \citenamefont {Hansen}}]{Jaffe:2005Evidence}%
  \BibitemOpen
  \bibfield  {author} {\bibinfo {author} {\bibfnamefont {T.~R.}\ \bibnamefont
  {Jaffe}}, \bibinfo {author} {\bibfnamefont {A.~J.}\ \bibnamefont {Banday}},
  \bibinfo {author} {\bibfnamefont {H.~K.}\ \bibnamefont {Eriksen}}, \bibinfo
  {author} {\bibfnamefont {K.~M.}\ \bibnamefont {Gorski}},\ and\ \bibinfo
  {author} {\bibfnamefont {F.~K.}\ \bibnamefont {Hansen}},\ }\bibfield  {title}
  {\bibinfo {title} {{Evidence of vorticity and shear at large angular scales
  in the WMAP data: A Violation of cosmological isotropy?}},\ }\href
  {https://doi.org/10.1086/444454} {\bibfield  {journal} {\bibinfo  {journal}
  {Astrophys. J. Lett.}\ }\textbf {\bibinfo {volume} {629}},\ \bibinfo {pages}
  {L1} (\bibinfo {year} {2005})}\BibitemShut {NoStop}%
\bibitem [{\citenamefont {Jaffe}\ \emph
  {et~al.}(2006{\natexlab{a}})\citenamefont {Jaffe}, \citenamefont {Banday},
  \citenamefont {Eriksen}, \citenamefont {Gorski},\ and\ \citenamefont
  {Hansen}}]{Jaffe:2006Bianchi}%
  \BibitemOpen
  \bibfield  {author} {\bibinfo {author} {\bibfnamefont {T.~R.}\ \bibnamefont
  {Jaffe}}, \bibinfo {author} {\bibfnamefont {A.~J.}\ \bibnamefont {Banday}},
  \bibinfo {author} {\bibfnamefont {H.~K.}\ \bibnamefont {Eriksen}}, \bibinfo
  {author} {\bibfnamefont {K.~M.}\ \bibnamefont {Gorski}},\ and\ \bibinfo
  {author} {\bibfnamefont {F.~K.}\ \bibnamefont {Hansen}},\ }\bibfield  {title}
  {\bibinfo {title} {{Bianchi Type VII(h) Models and the WMAP 3-year Data}},\
  }\href {https://doi.org/10.1051/0004-6361:20065748} {\bibfield  {journal}
  {\bibinfo  {journal} {Astron. Astrophys.}\ }\textbf {\bibinfo {volume}
  {460}},\ \bibinfo {pages} {393} (\bibinfo {year}
  {2006}{\natexlab{a}})}\BibitemShut {NoStop}%
\bibitem [{\citenamefont {Jaffe}\ \emph
  {et~al.}(2006{\natexlab{b}})\citenamefont {Jaffe}, \citenamefont {Banday},
  \citenamefont {Eriksen}, \citenamefont {Gorski},\ and\ \citenamefont
  {Hansen}}]{Jaffe:2006Fast}%
  \BibitemOpen
  \bibfield  {author} {\bibinfo {author} {\bibfnamefont {T.~R.}\ \bibnamefont
  {Jaffe}}, \bibinfo {author} {\bibfnamefont {A.~J.}\ \bibnamefont {Banday}},
  \bibinfo {author} {\bibfnamefont {H.~K.}\ \bibnamefont {Eriksen}}, \bibinfo
  {author} {\bibfnamefont {K.~M.}\ \bibnamefont {Gorski}},\ and\ \bibinfo
  {author} {\bibfnamefont {F.~K.}\ \bibnamefont {Hansen}},\ }\bibfield  {title}
  {\bibinfo {title} {{Fast and efficient template fitting of deterministic
  anisotropic cosmological models applied to wmap data}},\ }\href
  {https://doi.org/10.1086/501343} {\bibfield  {journal} {\bibinfo  {journal}
  {Astrophys. J.}\ }\textbf {\bibinfo {volume} {643}},\ \bibinfo {pages} {616}
  (\bibinfo {year} {2006}{\natexlab{b}})}\BibitemShut {NoStop}%
\bibitem [{\citenamefont {Campanelli}\ \emph {et~al.}(2006)\citenamefont
  {Campanelli}, \citenamefont {Cea},\ and\ \citenamefont
  {Tedesco}}]{Campanelli:2006Ellipsoidal}%
  \BibitemOpen
  \bibfield  {author} {\bibinfo {author} {\bibfnamefont {L.}~\bibnamefont
  {Campanelli}}, \bibinfo {author} {\bibfnamefont {P.}~\bibnamefont {Cea}},\
  and\ \bibinfo {author} {\bibfnamefont {L.}~\bibnamefont {Tedesco}},\
  }\bibfield  {title} {\bibinfo {title} {{Ellipsoidal Universe Can Solve The
  CMB Quadrupole Problem}},\ }\href
  {https://doi.org/10.1103/PhysRevLett.97.131302} {\bibfield  {journal}
  {\bibinfo  {journal} {Phys. Rev. Lett.}\ }\textbf {\bibinfo {volume} {97}},\
  \bibinfo {pages} {131302} (\bibinfo {year} {2006})},\ \bibinfo {note}
  {[Erratum: Phys.Rev.Lett. 97, 209903 (2006)]}\BibitemShut {NoStop}%
\bibitem [{\citenamefont {Campanelli}\ \emph {et~al.}(2007)\citenamefont
  {Campanelli}, \citenamefont {Cea},\ and\ \citenamefont
  {Tedesco}}]{Campanelli:2007Cosmic}%
  \BibitemOpen
  \bibfield  {author} {\bibinfo {author} {\bibfnamefont {L.}~\bibnamefont
  {Campanelli}}, \bibinfo {author} {\bibfnamefont {P.}~\bibnamefont {Cea}},\
  and\ \bibinfo {author} {\bibfnamefont {L.}~\bibnamefont {Tedesco}},\
  }\bibfield  {title} {\bibinfo {title} {{Cosmic Microwave Background
  Quadrupole and Ellipsoidal Universe}},\ }\href
  {https://doi.org/10.1103/PhysRevD.76.063007} {\bibfield  {journal} {\bibinfo
  {journal} {Phys. Rev. D}\ }\textbf {\bibinfo {volume} {76}},\ \bibinfo
  {pages} {063007} (\bibinfo {year} {2007})}\BibitemShut {NoStop}%
\bibitem [{\citenamefont {Vollick}(2003)}]{Vollick:2003AnisotropicBI}%
  \BibitemOpen
  \bibfield  {author} {\bibinfo {author} {\bibfnamefont {D.~N.}\ \bibnamefont
  {Vollick}},\ }\bibfield  {title} {\bibinfo {title} {{Anisotropic Born-Infeld
  cosmologies}},\ }\href {https://doi.org/10.1023/A:1024551105800} {\bibfield
  {journal} {\bibinfo  {journal} {Gen. Rel. Grav.}\ }\textbf {\bibinfo {volume}
  {35}},\ \bibinfo {pages} {1511} (\bibinfo {year} {2003})}\BibitemShut
  {NoStop}%
\bibitem [{\citenamefont {Ellis}(2006)}]{Ellis:2006TheBianchiModels}%
  \BibitemOpen
  \bibfield  {author} {\bibinfo {author} {\bibfnamefont {G.~F.~R.}\
  \bibnamefont {Ellis}},\ }\bibfield  {title} {\bibinfo {title} {{The Bianchi
  models: Then and now}},\ }\href {https://doi.org/10.1007/s10714-006-0283-4}
  {\bibfield  {journal} {\bibinfo  {journal} {Gen. Rel. Grav.}\ }\textbf
  {\bibinfo {volume} {38}},\ \bibinfo {pages} {1003} (\bibinfo {year}
  {2006})}\BibitemShut {NoStop}%
\bibitem [{\citenamefont {Saha}(2006)}]{Saha:2006Anisotropic}%
  \BibitemOpen
  \bibfield  {author} {\bibinfo {author} {\bibfnamefont {B.}~\bibnamefont
  {Saha}},\ }\bibfield  {title} {\bibinfo {title} {{Anisotropic cosmological
  models with a perfect fluid and a Lambda term}},\ }\href
  {https://doi.org/10.1007/s10509-005-9008-5} {\bibfield  {journal} {\bibinfo
  {journal} {Astrophys. Space Sci.}\ }\textbf {\bibinfo {volume} {302}},\
  \bibinfo {pages} {83} (\bibinfo {year} {2006})}\BibitemShut {NoStop}%
\bibitem [{\citenamefont {Gumrukcuoglu}\ \emph {et~al.}(2007)\citenamefont
  {Gumrukcuoglu}, \citenamefont {Contaldi},\ and\ \citenamefont
  {Peloso}}]{Gumrukcuoglu:2007Inflationary}%
  \BibitemOpen
  \bibfield  {author} {\bibinfo {author} {\bibfnamefont {A.~E.}\ \bibnamefont
  {Gumrukcuoglu}}, \bibinfo {author} {\bibfnamefont {C.~R.}\ \bibnamefont
  {Contaldi}},\ and\ \bibinfo {author} {\bibfnamefont {M.}~\bibnamefont
  {Peloso}},\ }\bibfield  {title} {\bibinfo {title} {{Inflationary
  perturbations in anisotropic backgrounds and their imprint on the CMB}},\
  }\href {https://doi.org/10.1088/1475-7516/2007/11/005} {\bibfield  {journal}
  {\bibinfo  {journal} {JCAP}\ }\textbf {\bibinfo {volume} {11}},\ \bibinfo
  {pages} {005}}\BibitemShut {NoStop}%
\bibitem [{\citenamefont {Rodrigues}(2008)}]{Rodrigues:2008Evolution}%
  \BibitemOpen
  \bibfield  {author} {\bibinfo {author} {\bibfnamefont {D.~C.}\ \bibnamefont
  {Rodrigues}},\ }\bibfield  {title} {\bibinfo {title} {{Evolution of
  Anisotropies in Eddington-Born-Infeld Cosmology}},\ }\href
  {https://doi.org/10.1103/PhysRevD.78.063013} {\bibfield  {journal} {\bibinfo
  {journal} {Phys. Rev. D}\ }\textbf {\bibinfo {volume} {78}},\ \bibinfo
  {pages} {063013} (\bibinfo {year} {2008})}\BibitemShut {NoStop}%
\bibitem [{\citenamefont {Akarsu}\ and\ \citenamefont
  {Kilinc}(2010)}]{Akarsu:2010Bianchi}%
  \BibitemOpen
  \bibfield  {author} {\bibinfo {author} {\bibfnamefont {O.}~\bibnamefont
  {Akarsu}}\ and\ \bibinfo {author} {\bibfnamefont {C.~B.}\ \bibnamefont
  {Kilinc}},\ }\bibfield  {title} {\bibinfo {title} {{Bianchi type III models
  with anisotropic dark energy}},\ }\href
  {https://doi.org/10.1007/s10714-009-0878-7} {\bibfield  {journal} {\bibinfo
  {journal} {Gen. Rel. Grav.}\ }\textbf {\bibinfo {volume} {42}},\ \bibinfo
  {pages} {763} (\bibinfo {year} {2010})}\BibitemShut {NoStop}%
\bibitem [{\citenamefont {Sharif}\ and\ \citenamefont
  {Kausar}(2011)}]{Sharif:2011Anisotropic}%
  \BibitemOpen
  \bibfield  {author} {\bibinfo {author} {\bibfnamefont {M.}~\bibnamefont
  {Sharif}}\ and\ \bibinfo {author} {\bibfnamefont {H.~R.}\ \bibnamefont
  {Kausar}},\ }\bibfield  {title} {\bibinfo {title} {{Anisotropic Fluid and
  Bianchi Type III Model in f(R) Gravity}},\ }\href
  {https://doi.org/10.1016/j.physletb.2011.01.027} {\bibfield  {journal}
  {\bibinfo  {journal} {Phys. Lett. B}\ }\textbf {\bibinfo {volume} {697}},\
  \bibinfo {pages} {1} (\bibinfo {year} {2011})}\BibitemShut {NoStop}%
\bibitem [{\citenamefont {Harko}\ \emph {et~al.}(2014)\citenamefont {Harko},
  \citenamefont {Lobo},\ and\ \citenamefont {Mak}}]{Harko:2014Bianchi}%
  \BibitemOpen
  \bibfield  {author} {\bibinfo {author} {\bibfnamefont {T.}~\bibnamefont
  {Harko}}, \bibinfo {author} {\bibfnamefont {F.~S.~N.}\ \bibnamefont {Lobo}},\
  and\ \bibinfo {author} {\bibfnamefont {M.~K.}\ \bibnamefont {Mak}},\
  }\bibfield  {title} {\bibinfo {title} {{Bianchi type I cosmological models in
  Eddington-inspired Born-Infeld gravity}},\ }\href
  {https://doi.org/10.3390/galaxies2040496} {\bibfield  {journal} {\bibinfo
  {journal} {Galaxies}\ }\textbf {\bibinfo {volume} {2}},\ \bibinfo {pages}
  {496} (\bibinfo {year} {2014})}\BibitemShut {NoStop}%
\bibitem [{\citenamefont {M\"uller}\ \emph {et~al.}(2018)\citenamefont
  {M\"uller}, \citenamefont {Ricciardone}, \citenamefont {Starobinsky},\ and\
  \citenamefont {Toporensky}}]{Muller:2018Anisotropic}%
  \BibitemOpen
  \bibfield  {author} {\bibinfo {author} {\bibfnamefont {D.}~\bibnamefont
  {M\"uller}}, \bibinfo {author} {\bibfnamefont {A.}~\bibnamefont
  {Ricciardone}}, \bibinfo {author} {\bibfnamefont {A.~A.}\ \bibnamefont
  {Starobinsky}},\ and\ \bibinfo {author} {\bibfnamefont {A.}~\bibnamefont
  {Toporensky}},\ }\bibfield  {title} {\bibinfo {title} {{Anisotropic
  cosmological solutions in $R + R^2$ gravity}},\ }\href
  {https://doi.org/10.1140/epjc/s10052-018-5778-0} {\bibfield  {journal}
  {\bibinfo  {journal} {Eur. Phys. J. C}\ }\textbf {\bibinfo {volume} {78}},\
  \bibinfo {pages} {311} (\bibinfo {year} {2018})},\ \Eprint
  {https://arxiv.org/abs/1710.08753} {arXiv:1710.08753 [gr-qc]} \BibitemShut
  {NoStop}%
\bibitem [{\citenamefont {Kumar}\ \emph {et~al.}(2021)\citenamefont {Kumar},
  \citenamefont {Maheshwari}, \citenamefont {Mazumdar},\ and\ \citenamefont
  {Peng}}]{Kumar:2021Anisotropic}%
  \BibitemOpen
  \bibfield  {author} {\bibinfo {author} {\bibfnamefont {K.~S.}\ \bibnamefont
  {Kumar}}, \bibinfo {author} {\bibfnamefont {S.}~\bibnamefont {Maheshwari}},
  \bibinfo {author} {\bibfnamefont {A.}~\bibnamefont {Mazumdar}},\ and\
  \bibinfo {author} {\bibfnamefont {J.}~\bibnamefont {Peng}},\ }\bibfield
  {title} {\bibinfo {title} {{An anisotropic bouncing universe in non-local
  gravity}},\ }\href {https://doi.org/10.1088/1475-7516/2021/07/025} {\bibfield
   {journal} {\bibinfo  {journal} {JCAP}\ }\textbf {\bibinfo {volume} {07}},\
  \bibinfo {pages} {025}}\BibitemShut {NoStop}%
\bibitem [{\citenamefont {Costantini}\ and\ \citenamefont
  {Elizalde}(2022)}]{Costantini:2022AReconstruction}%
  \BibitemOpen
  \bibfield  {author} {\bibinfo {author} {\bibfnamefont {A.}~\bibnamefont
  {Costantini}}\ and\ \bibinfo {author} {\bibfnamefont {E.}~\bibnamefont
  {Elizalde}},\ }\bibfield  {title} {\bibinfo {title} {{A reconstruction method
  for anisotropic universes in unimodular F(R)-gravity}},\ }\href
  {https://doi.org/10.1140/epjc/s10052-022-11112-3} {\bibfield  {journal}
  {\bibinfo  {journal} {Eur. Phys. J. C}\ }\textbf {\bibinfo {volume} {82}},\
  \bibinfo {pages} {1127} (\bibinfo {year} {2022})}\BibitemShut {NoStop}%
\bibitem [{\citenamefont {Nojiri}\ \emph
  {et~al.}(2022{\natexlab{a}})\citenamefont {Nojiri}, \citenamefont {Odintsov},
  \citenamefont {Oikonomou},\ and\ \citenamefont
  {Constantini}}]{Nojiri:2022Formalizing}%
  \BibitemOpen
  \bibfield  {author} {\bibinfo {author} {\bibfnamefont {S.}~\bibnamefont
  {Nojiri}}, \bibinfo {author} {\bibfnamefont {S.~D.}\ \bibnamefont
  {Odintsov}}, \bibinfo {author} {\bibfnamefont {V.~K.}\ \bibnamefont
  {Oikonomou}},\ and\ \bibinfo {author} {\bibfnamefont {A.}~\bibnamefont
  {Constantini}},\ }\bibfield  {title} {\bibinfo {title} {{Formalizing
  anisotropic inflation in modified gravity}},\ }\href
  {https://doi.org/10.1016/j.nuclphysb.2022.116011} {\bibfield  {journal}
  {\bibinfo  {journal} {Nucl. Phys. B}\ }\textbf {\bibinfo {volume} {985}},\
  \bibinfo {pages} {116011} (\bibinfo {year} {2022}{\natexlab{a}})}\BibitemShut
  {NoStop}%
\bibitem [{\citenamefont {Parnovsky}(2023)}]{Parnovsky:2023TheBigBang}%
  \BibitemOpen
  \bibfield  {author} {\bibinfo {author} {\bibfnamefont {S.~L.}\ \bibnamefont
  {Parnovsky}},\ }\bibfield  {title} {\bibinfo {title} {{The Big Bang could be
  anisotropic. The case of Bianchi I model}},\ }\href
  {https://doi.org/10.1088/1361-6382/acd7c2} {\bibfield  {journal} {\bibinfo
  {journal} {Class. Quant. Grav.}\ }\textbf {\bibinfo {volume} {40}},\ \bibinfo
  {pages} {135005} (\bibinfo {year} {2023})}\BibitemShut {NoStop}%
\bibitem [{\citenamefont {Kibaro\u{g}lu}(2025)}]{Kibaroglu:2025Anisotropic}%
  \BibitemOpen
  \bibfield  {author} {\bibinfo {author} {\bibfnamefont {S.}~\bibnamefont
  {Kibaro\u{g}lu}},\ }\bibfield  {title} {\bibinfo {title} {{Anisotropic
  Born-Infeld-f(R) cosmologies}},\ }\href
  {https://doi.org/10.1016/j.dark.2024.101784} {\bibfield  {journal} {\bibinfo
  {journal} {Phys. Dark Univ.}\ }\textbf {\bibinfo {volume} {47}},\ \bibinfo
  {pages} {101784} (\bibinfo {year} {2025})}\BibitemShut {NoStop}%
\bibitem [{\citenamefont {Parthasarathy}\ and\ \citenamefont
  {Viswanathan}(1991)}]{Parthasarathy:1991A-Q-Analog}%
  \BibitemOpen
  \bibfield  {author} {\bibinfo {author} {\bibfnamefont {R.}~\bibnamefont
  {Parthasarathy}}\ and\ \bibinfo {author} {\bibfnamefont {K.~S.}\ \bibnamefont
  {Viswanathan}},\ }\bibfield  {title} {\bibinfo {title} {{A q analog of the
  supersymmetric oscillator and its q superalgebra}},\ }\href
  {https://doi.org/10.1088/0305-4470/24/3/019} {\bibfield  {journal} {\bibinfo
  {journal} {J. Phys. A}\ }\textbf {\bibinfo {volume} {24}},\ \bibinfo {pages}
  {613} (\bibinfo {year} {1991})}\BibitemShut {NoStop}%
\bibitem [{\citenamefont {Viswanathan}\ \emph {et~al.}(1992)\citenamefont
  {Viswanathan}, \citenamefont {Parthasarathy},\ and\ \citenamefont
  {Jagannathan}}]{Viswanathan:1992Generalized}%
  \BibitemOpen
  \bibfield  {author} {\bibinfo {author} {\bibfnamefont {K.~S.}\ \bibnamefont
  {Viswanathan}}, \bibinfo {author} {\bibfnamefont {R.}~\bibnamefont
  {Parthasarathy}},\ and\ \bibinfo {author} {\bibfnamefont {R.}~\bibnamefont
  {Jagannathan}},\ }\bibfield  {title} {\bibinfo {title} {{Generalized q
  fermion oscillators and q coherent states}},\ }\href
  {https://doi.org/10.1088/0305-4470/25/7/009} {\bibfield  {journal} {\bibinfo
  {journal} {J. Phys. A}\ }\textbf {\bibinfo {volume} {25}},\ \bibinfo {pages}
  {L335} (\bibinfo {year} {1992})}\BibitemShut {NoStop}%
\bibitem [{\citenamefont {Chaichian}\ \emph {et~al.}(1993)\citenamefont
  {Chaichian}, \citenamefont {Gonzalez~Felipe},\ and\ \citenamefont
  {Montonen}}]{Chaichian:1993Statistics}%
  \BibitemOpen
  \bibfield  {author} {\bibinfo {author} {\bibfnamefont {M.}~\bibnamefont
  {Chaichian}}, \bibinfo {author} {\bibfnamefont {R.}~\bibnamefont
  {Gonzalez~Felipe}},\ and\ \bibinfo {author} {\bibfnamefont {C.}~\bibnamefont
  {Montonen}},\ }\bibfield  {title} {\bibinfo {title} {{Statistics of q
  oscillators, quons and relations to fractional statistics}},\ }\href
  {https://doi.org/10.1088/0305-4470/26/16/018} {\bibfield  {journal} {\bibinfo
   {journal} {J. Phys. A}\ }\textbf {\bibinfo {volume} {26}},\ \bibinfo {pages}
  {4017} (\bibinfo {year} {1993})}\BibitemShut {NoStop}%
\bibitem [{\citenamefont {Algin}\ and\ \citenamefont
  {Senay}(2012)}]{Algin:2012High}%
  \BibitemOpen
  \bibfield  {author} {\bibinfo {author} {\bibfnamefont {A.}~\bibnamefont
  {Algin}}\ and\ \bibinfo {author} {\bibfnamefont {M.}~\bibnamefont {Senay}},\
  }\bibfield  {title} {\bibinfo {title} {{High temperature behavior of a
  deformed Fermi gas obeying interpolating statistics}},\ }\href
  {https://doi.org/10.1103/PhysRevE.85.041123} {\bibfield  {journal} {\bibinfo
  {journal} {Phys. Rev. E}\ }\textbf {\bibinfo {volume} {85}},\ \bibinfo
  {pages} {041123} (\bibinfo {year} {2012})}\BibitemShut {NoStop}%
\bibitem [{\citenamefont {Algin}\ and\ \citenamefont
  {Senay}(2016{\natexlab{a}})}]{Algin:2016Fermionic}%
  \BibitemOpen
  \bibfield  {author} {\bibinfo {author} {\bibfnamefont {A.}~\bibnamefont
  {Algin}}\ and\ \bibinfo {author} {\bibfnamefont {M.}~\bibnamefont {Senay}},\
  }\bibfield  {title} {\bibinfo {title} {{Fermionic q -deformation and its
  connection to thermal effective mass of a quasiparticle}},\ }\href
  {https://doi.org/10.1016/j.physa.2015.12.014} {\bibfield  {journal} {\bibinfo
   {journal} {Physica A}\ }\textbf {\bibinfo {volume} {447}},\ \bibinfo {pages}
  {232} (\bibinfo {year} {2016}{\natexlab{a}})}\BibitemShut {NoStop}%
\bibitem [{\citenamefont {Algin}\ and\ \citenamefont
  {Senay}(2016{\natexlab{b}})}]{Algin:2016General}%
  \BibitemOpen
  \bibfield  {author} {\bibinfo {author} {\bibfnamefont {A.}~\bibnamefont
  {Algin}}\ and\ \bibinfo {author} {\bibfnamefont {M.}~\bibnamefont {Senay}},\
  }\bibfield  {title} {\bibinfo {title} {{General thermostatistical properties
  of a q-deformed fermion gas in two dimensions}},\ }\href
  {https://doi.org/10.1088/1742-6596/766/1/012008} {\bibfield  {journal}
  {\bibinfo  {journal} {J. Phys. Conf. Ser.}\ }\textbf {\bibinfo {volume}
  {766}},\ \bibinfo {pages} {012008} (\bibinfo {year}
  {2016}{\natexlab{b}})}\BibitemShut {NoStop}%
\bibitem [{\citenamefont {Verlinde}(2011)}]{Verlinde:2010hp}%
  \BibitemOpen
  \bibfield  {author} {\bibinfo {author} {\bibfnamefont {E.~P.}\ \bibnamefont
  {Verlinde}},\ }\bibfield  {title} {\bibinfo {title} {{On the Origin of
  Gravity and the Laws of Newton}},\ }\href
  {https://doi.org/10.1007/JHEP04(2011)029} {\bibfield  {journal} {\bibinfo
  {journal} {JHEP}\ }\textbf {\bibinfo {volume} {04}},\ \bibinfo {pages}
  {029}}\BibitemShut {NoStop}%
\bibitem [{\citenamefont {Verlinde}(2017)}]{Verlinde:2016toy}%
  \BibitemOpen
  \bibfield  {author} {\bibinfo {author} {\bibfnamefont {E.~P.}\ \bibnamefont
  {Verlinde}},\ }\bibfield  {title} {\bibinfo {title} {{Emergent Gravity and
  the Dark Universe}},\ }\href {https://doi.org/10.21468/SciPostPhys.2.3.016}
  {\bibfield  {journal} {\bibinfo  {journal} {SciPost Phys.}\ }\textbf
  {\bibinfo {volume} {2}},\ \bibinfo {pages} {016} (\bibinfo {year}
  {2017})}\BibitemShut {NoStop}%
\bibitem [{\citenamefont {Unruh}(1976)}]{Unruh:1976Notes}%
  \BibitemOpen
  \bibfield  {author} {\bibinfo {author} {\bibfnamefont {W.~G.}\ \bibnamefont
  {Unruh}},\ }\bibfield  {title} {\bibinfo {title} {{Notes on black hole
  evaporation}},\ }\href {https://doi.org/10.1103/PhysRevD.14.870} {\bibfield
  {journal} {\bibinfo  {journal} {Phys. Rev. D}\ }\textbf {\bibinfo {volume}
  {14}},\ \bibinfo {pages} {870} (\bibinfo {year} {1976})}\BibitemShut
  {NoStop}%
\bibitem [{\citenamefont {Wald}(2010)}]{Wald:2010GeneralRelativity}%
  \BibitemOpen
  \bibfield  {author} {\bibinfo {author} {\bibfnamefont {R.~M.}\ \bibnamefont
  {Wald}},\ }\href@noop {} {\emph {\bibinfo {title} {General relativity}}}\
  (\bibinfo  {publisher} {University of Chicago press},\ \bibinfo {year}
  {2010})\BibitemShut {NoStop}%
\bibitem [{\citenamefont {Taub}(1951)}]{Taub:1951Empty}%
  \BibitemOpen
  \bibfield  {author} {\bibinfo {author} {\bibfnamefont {A.~H.}\ \bibnamefont
  {Taub}},\ }\bibfield  {title} {\bibinfo {title} {{Empty space-times admitting
  a three parameter group of motions}},\ }\href
  {https://doi.org/10.2307/1969567} {\bibfield  {journal} {\bibinfo  {journal}
  {Annals Math.}\ }\textbf {\bibinfo {volume} {53}},\ \bibinfo {pages} {472}
  (\bibinfo {year} {1951})}\BibitemShut {NoStop}%
\bibitem [{\citenamefont {Barrow}\ and\ \citenamefont
  {Clifton}(2006)}]{Barrow:2006Cosmologies}%
  \BibitemOpen
  \bibfield  {author} {\bibinfo {author} {\bibfnamefont {J.~D.}\ \bibnamefont
  {Barrow}}\ and\ \bibinfo {author} {\bibfnamefont {T.}~\bibnamefont
  {Clifton}},\ }\bibfield  {title} {\bibinfo {title} {{Cosmologies with energy
  exchange}},\ }\href {https://doi.org/10.1103/PhysRevD.73.103520} {\bibfield
  {journal} {\bibinfo  {journal} {Phys. Rev. D}\ }\textbf {\bibinfo {volume}
  {73}},\ \bibinfo {pages} {103520} (\bibinfo {year} {2006})}\BibitemShut
  {NoStop}%
\bibitem [{\citenamefont {Berera}\ \emph {et~al.}(2004)\citenamefont {Berera},
  \citenamefont {Buniy},\ and\ \citenamefont
  {Kephart}}]{Berera:2004TheEccentric}%
  \BibitemOpen
  \bibfield  {author} {\bibinfo {author} {\bibfnamefont {A.}~\bibnamefont
  {Berera}}, \bibinfo {author} {\bibfnamefont {R.~V.}\ \bibnamefont {Buniy}},\
  and\ \bibinfo {author} {\bibfnamefont {T.~W.}\ \bibnamefont {Kephart}},\
  }\bibfield  {title} {\bibinfo {title} {{The Eccentric universe}},\ }\href
  {https://doi.org/10.1088/1475-7516/2004/10/016} {\bibfield  {journal}
  {\bibinfo  {journal} {JCAP}\ }\textbf {\bibinfo {volume} {10}},\ \bibinfo
  {pages} {016}}\BibitemShut {NoStop}%
\bibitem [{\citenamefont {Campanelli}(2009)}]{Campanelli:2009AModel}%
  \BibitemOpen
  \bibfield  {author} {\bibinfo {author} {\bibfnamefont {L.}~\bibnamefont
  {Campanelli}},\ }\bibfield  {title} {\bibinfo {title} {{A Model of Universe
  Anisotropization}},\ }\href {https://doi.org/10.1103/PhysRevD.80.063006}
  {\bibfield  {journal} {\bibinfo  {journal} {Phys. Rev. D}\ }\textbf {\bibinfo
  {volume} {80}},\ \bibinfo {pages} {063006} (\bibinfo {year}
  {2009})}\BibitemShut {NoStop}%
\bibitem [{\citenamefont {Deliduman}\ \emph {et~al.}(2024)\citenamefont
  {Deliduman}, \citenamefont {Kasikci},\ and\ \citenamefont
  {Tugyanoglu}}]{Deliduman:2024ellipsoidal}%
  \BibitemOpen
  \bibfield  {author} {\bibinfo {author} {\bibfnamefont {C.}~\bibnamefont
  {Deliduman}}, \bibinfo {author} {\bibfnamefont {O.}~\bibnamefont {Kasikci}},\
  and\ \bibinfo {author} {\bibfnamefont {V.~K.}\ \bibnamefont {Tugyanoglu}},\
  }\bibfield  {title} {\bibinfo {title} {{f(R) gravity in an ellipsoidal
  universe}},\ }\href {https://doi.org/10.1016/j.dark.2024.101469} {\bibfield
  {journal} {\bibinfo  {journal} {Phys. Dark Univ.}\ }\textbf {\bibinfo
  {volume} {44}},\ \bibinfo {pages} {101469} (\bibinfo {year} {2024})},\
  \Eprint {https://arxiv.org/abs/2310.02914} {arXiv:2310.02914 [gr-qc]}
  \BibitemShut {NoStop}%
\bibitem [{\citenamefont {Peter}\ and\ \citenamefont
  {Pinto-Neto}(2002)}]{Peter:2002Primordial}%
  \BibitemOpen
  \bibfield  {author} {\bibinfo {author} {\bibfnamefont {P.}~\bibnamefont
  {Peter}}\ and\ \bibinfo {author} {\bibfnamefont {N.}~\bibnamefont
  {Pinto-Neto}},\ }\bibfield  {title} {\bibinfo {title} {{Primordial
  perturbations in a non singular bouncing universe model}},\ }\href
  {https://doi.org/10.1103/PhysRevD.66.063509} {\bibfield  {journal} {\bibinfo
  {journal} {Phys. Rev. D}\ }\textbf {\bibinfo {volume} {66}},\ \bibinfo
  {pages} {063509} (\bibinfo {year} {2002})}\BibitemShut {NoStop}%
\bibitem [{\citenamefont {Allen}\ and\ \citenamefont
  {Wands}(2004)}]{Allen:2004Cosmological}%
  \BibitemOpen
  \bibfield  {author} {\bibinfo {author} {\bibfnamefont {L.~E.}\ \bibnamefont
  {Allen}}\ and\ \bibinfo {author} {\bibfnamefont {D.}~\bibnamefont {Wands}},\
  }\bibfield  {title} {\bibinfo {title} {{Cosmological perturbations through a
  simple bounce}},\ }\href {https://doi.org/10.1103/PhysRevD.70.063515}
  {\bibfield  {journal} {\bibinfo  {journal} {Phys. Rev. D}\ }\textbf {\bibinfo
  {volume} {70}},\ \bibinfo {pages} {063515} (\bibinfo {year}
  {2004})}\BibitemShut {NoStop}%
\bibitem [{\citenamefont {Cai}\ \emph {et~al.}(2011)\citenamefont {Cai},
  \citenamefont {Chen}, \citenamefont {Dent}, \citenamefont {Dutta},\ and\
  \citenamefont {Saridakis}}]{Cai:2011MatterBounce}%
  \BibitemOpen
  \bibfield  {author} {\bibinfo {author} {\bibfnamefont {Y.-F.}\ \bibnamefont
  {Cai}}, \bibinfo {author} {\bibfnamefont {S.-H.}\ \bibnamefont {Chen}},
  \bibinfo {author} {\bibfnamefont {J.~B.}\ \bibnamefont {Dent}}, \bibinfo
  {author} {\bibfnamefont {S.}~\bibnamefont {Dutta}},\ and\ \bibinfo {author}
  {\bibfnamefont {E.~N.}\ \bibnamefont {Saridakis}},\ }\bibfield  {title}
  {\bibinfo {title} {{Matter Bounce Cosmology with the f(T) Gravity}},\ }\href
  {https://doi.org/10.1088/0264-9381/28/21/215011} {\bibfield  {journal}
  {\bibinfo  {journal} {Class. Quant. Grav.}\ }\textbf {\bibinfo {volume}
  {28}},\ \bibinfo {pages} {215011} (\bibinfo {year} {2011})}\BibitemShut
  {NoStop}%
\bibitem [{\citenamefont {Cai}\ \emph {et~al.}(2012)\citenamefont {Cai},
  \citenamefont {Easson},\ and\ \citenamefont
  {Brandenberger}}]{Cai:2012Towards}%
  \BibitemOpen
  \bibfield  {author} {\bibinfo {author} {\bibfnamefont {Y.-F.}\ \bibnamefont
  {Cai}}, \bibinfo {author} {\bibfnamefont {D.~A.}\ \bibnamefont {Easson}},\
  and\ \bibinfo {author} {\bibfnamefont {R.}~\bibnamefont {Brandenberger}},\
  }\bibfield  {title} {\bibinfo {title} {{Towards a Nonsingular Bouncing
  Cosmology}},\ }\href {https://doi.org/10.1088/1475-7516/2012/08/020}
  {\bibfield  {journal} {\bibinfo  {journal} {JCAP}\ }\textbf {\bibinfo
  {volume} {08}},\ \bibinfo {pages} {020}},\ \Eprint
  {https://arxiv.org/abs/1206.2382} {arXiv:1206.2382 [hep-th]} \BibitemShut
  {NoStop}%
\bibitem [{\citenamefont {Nojiri}\ \emph {et~al.}(2019)\citenamefont {Nojiri},
  \citenamefont {Odintsov}, \citenamefont {Oikonomou},\ and\ \citenamefont
  {Paul}}]{Nojiri:2019NonsingularBounce}%
  \BibitemOpen
  \bibfield  {author} {\bibinfo {author} {\bibfnamefont {S.}~\bibnamefont
  {Nojiri}}, \bibinfo {author} {\bibfnamefont {S.~D.}\ \bibnamefont
  {Odintsov}}, \bibinfo {author} {\bibfnamefont {V.~K.}\ \bibnamefont
  {Oikonomou}},\ and\ \bibinfo {author} {\bibfnamefont {T.}~\bibnamefont
  {Paul}},\ }\bibfield  {title} {\bibinfo {title} {{Nonsingular bounce
  cosmology from Lagrange multiplier $F(R)$ gravity}},\ }\href
  {https://doi.org/10.1103/PhysRevD.100.084056} {\bibfield  {journal} {\bibinfo
   {journal} {Phys. Rev. D}\ }\textbf {\bibinfo {volume} {100}},\ \bibinfo
  {pages} {084056} (\bibinfo {year} {2019})}\BibitemShut {NoStop}%
\bibitem [{\citenamefont {Odintsov}\ \emph {et~al.}(2020)\citenamefont
  {Odintsov}, \citenamefont {Oikonomou},\ and\ \citenamefont
  {Paul}}]{Odintsov:2020FromBounce}%
  \BibitemOpen
  \bibfield  {author} {\bibinfo {author} {\bibfnamefont {S.~D.}\ \bibnamefont
  {Odintsov}}, \bibinfo {author} {\bibfnamefont {V.~K.}\ \bibnamefont
  {Oikonomou}},\ and\ \bibinfo {author} {\bibfnamefont {T.}~\bibnamefont
  {Paul}},\ }\bibfield  {title} {\bibinfo {title} {{From a Bounce to the Dark
  Energy Era with $F(R)$ Gravity}},\ }\href
  {https://doi.org/10.1088/1361-6382/abbc47} {\bibfield  {journal} {\bibinfo
  {journal} {Class. Quant. Grav.}\ }\textbf {\bibinfo {volume} {37}},\ \bibinfo
  {pages} {235005} (\bibinfo {year} {2020})}\BibitemShut {NoStop}%
\bibitem [{\citenamefont {Bernal}\ \emph {et~al.}(2016)\citenamefont {Bernal},
  \citenamefont {Verde},\ and\ \citenamefont
  {Riess}}]{Bernal:2016TheTroubleH0}%
  \BibitemOpen
  \bibfield  {author} {\bibinfo {author} {\bibfnamefont {J.~L.}\ \bibnamefont
  {Bernal}}, \bibinfo {author} {\bibfnamefont {L.}~\bibnamefont {Verde}},\ and\
  \bibinfo {author} {\bibfnamefont {A.~G.}\ \bibnamefont {Riess}},\ }\bibfield
  {title} {\bibinfo {title} {{The trouble with $H_0$}},\ }\href
  {https://doi.org/10.1088/1475-7516/2016/10/019} {\bibfield  {journal}
  {\bibinfo  {journal} {JCAP}\ }\textbf {\bibinfo {volume} {10}},\ \bibinfo
  {pages} {019}}\BibitemShut {NoStop}%
\bibitem [{\citenamefont {Vagnozzi}(2023)}]{Vagnozzi:2023Seven}%
  \BibitemOpen
  \bibfield  {author} {\bibinfo {author} {\bibfnamefont {S.}~\bibnamefont
  {Vagnozzi}},\ }\bibfield  {title} {\bibinfo {title} {{Seven Hints That
  Early-Time New Physics Alone Is Not Sufficient to Solve the Hubble
  Tension}},\ }\href {https://doi.org/10.3390/universe9090393} {\bibfield
  {journal} {\bibinfo  {journal} {Universe}\ }\textbf {\bibinfo {volume} {9}},\
  \bibinfo {pages} {393} (\bibinfo {year} {2023})}\BibitemShut {NoStop}%
\bibitem [{\citenamefont {Nojiri}\ \emph
  {et~al.}(2022{\natexlab{b}})\citenamefont {Nojiri}, \citenamefont
  {Odintsov},\ and\ \citenamefont {Paul}}]{Nojiri:2022Early}%
  \BibitemOpen
  \bibfield  {author} {\bibinfo {author} {\bibfnamefont {S.}~\bibnamefont
  {Nojiri}}, \bibinfo {author} {\bibfnamefont {S.~D.}\ \bibnamefont
  {Odintsov}},\ and\ \bibinfo {author} {\bibfnamefont {T.}~\bibnamefont
  {Paul}},\ }\bibfield  {title} {\bibinfo {title} {{Early and late universe
  holographic cosmology from a new generalized entropy}},\ }\href
  {https://doi.org/10.1016/j.physletb.2022.137189} {\bibfield  {journal}
  {\bibinfo  {journal} {Phys. Lett. B}\ }\textbf {\bibinfo {volume} {831}},\
  \bibinfo {pages} {137189} (\bibinfo {year} {2022}{\natexlab{b}})}\BibitemShut
  {NoStop}%
\bibitem [{\citenamefont {Nojiri}\ \emph
  {et~al.}(2022{\natexlab{c}})\citenamefont {Nojiri}, \citenamefont
  {Odintsov},\ and\ \citenamefont {Paul}}]{Nojiri:2022Modified}%
  \BibitemOpen
  \bibfield  {author} {\bibinfo {author} {\bibfnamefont {S.}~\bibnamefont
  {Nojiri}}, \bibinfo {author} {\bibfnamefont {S.~D.}\ \bibnamefont
  {Odintsov}},\ and\ \bibinfo {author} {\bibfnamefont {T.}~\bibnamefont
  {Paul}},\ }\bibfield  {title} {\bibinfo {title} {{Modified cosmology from the
  thermodynamics of apparent horizon}},\ }\href
  {https://doi.org/10.1016/j.physletb.2022.137553} {\bibfield  {journal}
  {\bibinfo  {journal} {Phys. Lett. B}\ }\textbf {\bibinfo {volume} {835}},\
  \bibinfo {pages} {137553} (\bibinfo {year} {2022}{\natexlab{c}})}\BibitemShut
  {NoStop}%
\bibitem [{\citenamefont {Nojiri}\ \emph {et~al.}(2024)\citenamefont {Nojiri},
  \citenamefont {Odintsov},\ and\ \citenamefont {Paul}}]{Nojiri:2024Different}%
  \BibitemOpen
  \bibfield  {author} {\bibinfo {author} {\bibfnamefont {S.}~\bibnamefont
  {Nojiri}}, \bibinfo {author} {\bibfnamefont {S.~D.}\ \bibnamefont
  {Odintsov}},\ and\ \bibinfo {author} {\bibfnamefont {T.}~\bibnamefont
  {Paul}},\ }\bibfield  {title} {\bibinfo {title} {{Different Aspects of
  Entropic Cosmology}},\ }\href {https://doi.org/10.3390/universe10090352}
  {\bibfield  {journal} {\bibinfo  {journal} {Universe}\ }\textbf {\bibinfo
  {volume} {10}},\ \bibinfo {pages} {352} (\bibinfo {year} {2024})}\BibitemShut
  {NoStop}%
\end{thebibliography}%

\end{document}